# High-resolution timing electronics for fast pixel sensors

**Adriano Lai**[a,1] **and Gian-Matteo Cossu**[a,b]

[a]*Istituto Nazionale Fisica Nucleare, Sezione di Cagliari, Cagliari, Italy*
[b]*Dipartimento di Fisica, Università di Cagliari, Cagliari, Italy*

*E-mail:* adriano.lai@ca.infn.it

Abstract: Detectors based on pixels with timing capabilities are gaining increasing importance in the last years. Next-to-come high-energy physics experiments at colliders requires the use of time information in tracking, due to the increasing levels of track densities in the foreseen experimental conditions. Various different developments are ongoing on solid state sensors to gain high-resolution performance at the sensor level, as for example LGAD sensors or 3D sensors. Intrinsic sensor time resolution around 20 ps have been recently obtained. The increasing performance on the sensor side strongly demands an adequate development on the front-end electronics side, which now risks to become the performance bottle-neck in a tracking or vertex-detecting system. This paper aims to analyze the ultimate possible performance in timing of a typically-used front-end circuit, the Trans-Impedance Amplifier, considering different possible circuit configurations. Evidence to the preferable modes of operation in sensor read-out for timing measurement will be given.

---

[1]Corresponding author.

# Contents

## 1 Introduction

An important emerging requirement in experimental high energy physics concerns the need of introducing time measurements at the level of the single pixel sensor. As an example, the Ugrade-II of the LHCb experiment at the CERN LHC, scheduled to take data in about a decade from now, has requirements of concurrent space and time resolutions of the order of 10 $\mu$m and at least 50 ps respectively at the single pixel level [1]. Such a trend is foreseen to continue with more severe requirements in the subsequent generation of collider experiments [2], where time resolutions in the range of 10-20 ps per hit will be necessary. This poses a number of different technical issues both on the sensor and the front-end electronics side. Recent developments in fast silicon sensors [3, 4] demonstrate that time resolutions of 30-20 ps are already reachable on the sensor side. As a consequence, front-end electronics becomes a decisive limiting factor to high time resolution. In order not to degrade the sensor performance, the time resolution requirement for the front-end stage is having an electronic time jitter below 10 ps r.m.s, which is not trivial to obtain.

The problem of ps-fast front-end electronics can be attacked from different sides and points of view. A first perspective could concern the distinction among different circuit solutions and input stages. A second one concerns the technology choice (for example CMOS versus Si-Ge bipolar or BiCMOS, having superior intrinsic performance in terms of speed, but other limitations, as for example lower integration capabilities). A third one is about the problem of obtainable timing performance within limited area and power budget. This last perspective is particularly important about the development of pixels in vertex detectors with timing, where besides the input stage also a high precision Time-to-Digital-Converter has to be integrated.

The design of integrated pixel electronics for high resolution deserves a dedicated treatment and will be the subject of a separate work. The present paper is dedicated to explore the main requirements on fast input stages from the circuit scheme and characteristics point of view. In particular, the relationship and interaction between the characteristics of sensor and electronics are studied. This is important to define a clear path between the sensor operation and performance and the front-end design. Some simulation results on specific cases and specific values of sensor and electronics design parameters are also given, so to gain evidence of the impact of the various solutions explored on the timing performance of the System. Here and in the following the term System (with capital "S") will be regularly referred to sensor and front-end electronics, coupled together as a unique device.

## 2 Characteristics of solid state sensors and timing

A solid state pixel sensor or, more commonly, a pixel, can be considered as a capacitive sensor where the charge generated by ionization in the sensitive volume is collected at the electrodes by means of a suitable read-out circuit. The starting process, however, is the generation of current signals by induction, due to the movement of the charge carriers of both signs, made free by the ionizing tracks. In no-timing applications, where only the amount of collected charge is of interest, it is common practise to assimilate such current signals to delta-shaped pulses of infinitesimal duration. This is perfectly justified by the common use of relatively slow charge-integrating front-end stages, which make the charge collection time generally negligible. On the other hand, when timing issues are concerned, the fine structure of signals induced at the electrodes is crucial to understand and decide the final performance of the System. In particular, when design efforts on the sensor side provide devices with charge collection times at the deep sub-ns level, front-end performance should be able to exploit and not lose such an advantage.

When quoting the contributions to uncertainty in the measurement of time, the following main quantities are normally considered:

$$\sigma_t = \sqrt{\sigma_{tw}^2 + \sigma_{dr}^2 + \sigma_{TDC}^2 + \sigma_{un}^2 + \sigma_{ej}^2}, \qquad (2.1)$$

where $\sigma_\text{tw}$ (*time-walk*) depends on fluctuations of the signal amplitude, which cannot be minimised by design but only by dedicated signal processing; $\sigma_\text{dr}$ (*delta-rays*) depends on the effect of longitudinally disuniformities in the energy deposit due to delta rays (Landau tail); $\sigma_\text{TDC}$ depends on the digital resolution of the electronics (conversion error). The $\sigma_\text{un}$ contribution corresponds to the time dispersion caused by unevenness in the signal shapes, which are due to the different possible drift paths of the charge carriers in the sensor. The $\sigma_\text{un}$ term depends only on the geometry of the sensitive volume. In order to minimise the $\sigma_\text{un}$ term, maximum uniformity in the electric field must be obtained by design [5]. The $\sigma_\text{ej}$ (electronic jitter) term depends on the front-end electronics rise time and signal-to-noise ratio.

In the present work, we will not be interested on the effect of the $\sigma_\text{tw}$ and $\sigma_\text{TDC}$ terms. $\sigma_\text{tw}$ is considered a systematic uncertainty, which can be corrected by dedicated signal processing techniques. Similarly, $\sigma_\text{TDC}$ depends on the precision in time-to-digital conversion and not on the front-end circuit.

The $\sigma_\text{dr}$ and $\sigma_\text{un}$ terms decide the intrinsic sensor speed. Recent studies demonstrate that average time in charge collection distributions of 2-300 ps and standard deviations of 50-40 ps or less can be obtained [4, 5]. Here we will not consider this aspect of the matter in any specific detail and will tend to treat time distribution parameters of the sensors as System inputs. We will instead focus the analysis on the $\sigma_\text{ej}$ term, aiming to study the interaction of the front-end specifications and final characteristics with the sensor behavior and performance. The aim is to understand which is the optimal front-end to be designed for a given fast-timing sensor.

## 3  Front-end electronics for timing: the Trans-Impedance-Amplifier (TIA)

The traditional textbook solution for the read-out of capacitive sensors is the well-known Charge Sensitive Amplifier (CSA), possibly followed by a suitable number of differentiating (CR) and integrating (RC) stages, realising a so-called *Shaper* [6].

Actually, the CSA circuit is a particular case of a more general configuration, that is the Trans-Impedance-Amplifier (TIA) with shunt-shunt feedback (FB-TIA), schematically shown if figure 1 (left). In this circuit, a fraction of the voltage is taken at the output of the inverting amplifier and is converted to a current by the impedance of the feedback path that is subtracted from the input. This technique has the effect of lowering the input impedance of the amplifier (Miller effect) leading to a circuit that integrates the input signal. In the ideal case the input capacitance is $\approx AC_f$ and is big enough to make the system independent of the detector capacitance giving the output voltage:

$$V_{out} = Q_{in}/C_f \tag{3.1}$$

A simplified implementation of the TIA amplifier can be realized by a common-source NMOS in a so-called self-biased topology (fig 1, right).

The ideal CSA behavior is achieved if the feedback resistance $R_f$ and load resistance $R_D$ allow a high open-loop gain $A$ and an input impedance seen by the current generator that is given by:

$$Z_{in} \sim \frac{1}{sC_D} \parallel \frac{R_f}{A(1+s\tau_f)} \tag{3.2}$$

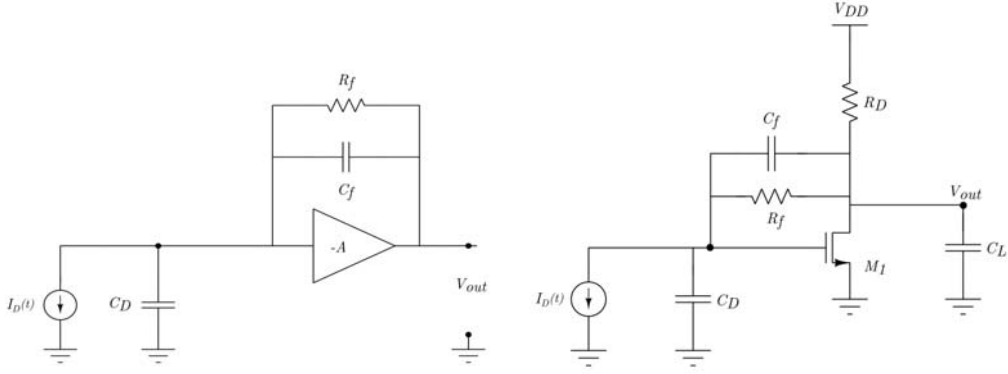

**Figure 1**. General representation of the FB-TIA circuit (left). NMOS FB-TIA with a self-biased topology (right). The current generator $I_D(t)$ and the $C_D$ capacitance model the operation of the capacitive sensor.

where $\tau_f = R_f C_f$. if $R_f \longrightarrow \infty$ and $AC_f \gg C_D$, then we have

$$Z_{in} \sim \frac{1}{sAC_f} \tag{3.3}$$

If the input current is considered as a Dirac Delta we have the input voltage:

$$V_{in} \sim -\frac{Q_{in}}{AC_f} \tag{3.4}$$

and the output voltage:

$$V_{out} = -AV_{in} = \frac{Q_{in}}{C_f} \tag{3.5}$$

Equation 3.5 describes the behavior of an ideal CSA, with the assumption of an infinitely fast amplifier (infinite bandwidth and slew-rate). A more realistic description can be obtain considering the small signal model of the circuit, as given in figure 2.

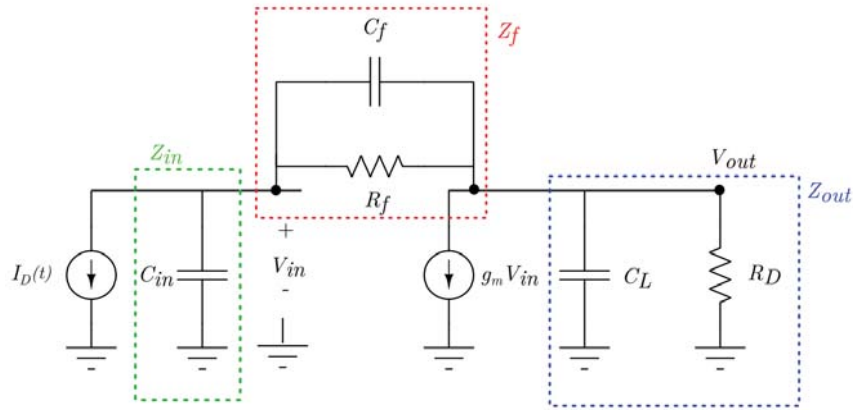

**Figure 2**. Small signal model of the TIA

The input capacitance is given by the sum of the detector capacitance and the one from gate and source of the NMOS transistor ($C_{in} = C_D + C_{gs}$). The output capacitance $C_L$ represent the total capacitance seen from the node $V_{out}$ to ground with open loop configuration and is given by the capacitance $C_{ds}$ and the input capacitance of the following stage. The resistance $R_D$ defines the gain of the common source which in given by $A \sim -g_m R_D$, where $g_m$ is the NMOS trans-conductance. We can define the impedances (see figure 2):

$$Z_{in} = \frac{1}{sC_{in}} \quad Z_f = \frac{R_f}{1+s\tau_f} \quad Z_{out} = \frac{R_D}{1+s\tau_D} \quad (3.6)$$

with the time constants: $\tau_f = R_f C_f$ ed $\tau_D = R_D C_L$. From the output node we find the voltage gain $G_v(s)$:

$$\frac{V_{in} - V_{out}}{Z_f} = g_m V_{in} + \frac{V_{out}}{Z_{out}} \quad (3.7)$$

and consequently:

$$G_v(s) = \frac{V_{out}}{V_{in}} = \frac{Z_{out}(1 - g_m Z_f)}{Z_{out} + Z_f}. \quad (3.8)$$

By replacing back the impedances, we obtain:

$$G_v(s) = \frac{R_D(1 + s\tau_f - g_m R_f)}{R_f(1 + s\tau_D) + R_D(1 + s\tau_f)}. \quad (3.9)$$

Introducing the parallel $R^* = R_f \parallel R_D$, the gain factor $G_0 = (g_m R^* - \frac{R^*}{R_f})$ and the time constants $\tau_f^* = R^* C_f$ and $\tau_L^* = R^*(C_f + C_L)$ the voltage gain can be written as:

$$G_v(s) = -\frac{G_0 - s\tau_f^*}{1 + s\tau_L^*} \quad (3.10)$$

The expression of $G_v(s)$ shows a pole at frequency $f_p = \frac{1}{2\pi \tau_L^*}$ and a zero at $f_z = \frac{G_0}{2\pi \tau_f^*} \approx \frac{g_m}{2\pi C_f}$. We can find the input impedance from the current equation at the input node:

$$I_D + \frac{V_{in}}{Z_{in}} + \frac{V_{in} - V_{out}}{Z_f} = 0. \quad (3.11)$$

Using equation 3.10 we get:

$$I_D = -V_{in}\left(\frac{1}{Z_{in}} + \frac{1}{Z_f}(1 - G_v)\right). \quad (3.12)$$

The voltage at the input of the circuit is then given by:

$$V_{in} = -I_D \frac{R_f(1 + s\tau_L^*)}{1 + G_0 + s(R_f(C_{in} + C_f(1+G_0)) + R^* C_L) + s^2 R^* R_f \xi} \quad (3.13)$$

with $\xi = C_L C_{in} + C_L C_f + C_{in} C_f$. The input impedance is by definition:

$$Z_{in_f} = \frac{V_{in}}{I_D} = \frac{R_f(1 + s\tau_L^*)}{1 + G_0 + s(R_f(C_{in} + C_f(1+G_0)) + R^* C_L) + s^2 R^* R_f \xi}, \quad (3.14)$$

from where the DC value of the input impedance can be derived, setting $s = 0$:

$$Z_{in_f} = \frac{R_f}{1+G_0}. \tag{3.15}$$

The output voltage is the product of the input voltage $V_{in}$ by the voltage gain $G_v(s)$. The product cancels out the zero in $V_{in}$ (equation 3.13) but introduces the zero of the voltage gain $G_v(s)$ (equation 3.10):

$$V_{out} = I_D \frac{R_f(G_0 - s\tau_f^*)}{1 + G_0 + s(R_f(C_{in} + C_f(1+G_0)) + R^*C_L) + s^2 R^* R_f \xi} \tag{3.16}$$

the trans-impedance of the circuit reads now:

$$\frac{V_{out}}{I_D} = \frac{R_f(G_0 - s\tau_f^*)}{1 + G_0 + s(R_f(C_{in} + C_f(1+G_0)) + R^*C_L) + s^2 R^* R_f \xi} \tag{3.17}$$

This important formula can be rearranged to introduce the natural frequency of the circuit $\omega_n$ and the damping factor $\zeta$ (see also [7]). Ignoring the zero for now, we can write the trans-impedance $R_m(s)$ as:

$$R_m(s) = \frac{K}{s^2 + 2\zeta\omega_n s + \omega_n^2} \tag{3.18}$$

where $K = R_f G_0/(R_f R^* \xi)$. The natural frequency is then given by:

$$\omega_n = \sqrt{\frac{1+G_0}{R^* R_f \xi}} \tag{3.19}$$

since $1 + G_0 \approx g_m R^*$, it follows that:

$$\omega_n \approx \sqrt{\frac{g_m}{R_f(C_L C_{in} + C_L C_f + C_{in} C_f)}} \tag{3.20}$$

which results independent of the load resistance $R_D$. The expression of the damping factor $\zeta$ is:

$$\zeta = \frac{1}{2}\frac{(R_f(C_{in} + C_f(1+G_0)) + R^* C_L)}{\sqrt{(1+G_0)R^* R_f \xi}}. \tag{3.21}$$

In general, if the damping factor $0 < \zeta < 1$, the poles are complex conjugated and we have an under-damped system, which can lead to an oscillating behavior. This condition is therefore to be avoided. On the other hand, if $\zeta \gg 1$, we get real and distinct poles and an over-dumped system [7]. Usually the system should be operated in a dumped condition, which is obtained at $\zeta \geq 1$.

In order to simplify our discussion, in the following we choose the circuit components and DC operating point in such a way to realize a *critically dumped* system ($\zeta = 1$). In this way, we obtain from equation 3.18 a second order transfer function with a unique negative pole:

$$R_m(s) = \frac{K}{(s+\omega_n)^2}. \tag{3.22}$$

Introducing the time constant:

$$\tau = \tau_n = \frac{1}{\omega_n} \tag{3.23}$$

and simplifying as follows (being $G_0 >> 1$):

$$\frac{K}{\omega_n^2} = \frac{R_f G_0}{R_f R^* \xi} \left( \frac{R_f R^* \xi}{1 + G_0} \right) \approx R_f, \quad (3.24)$$

we can write the transfer function of our TIA circuit as:

$$R_m(s) \approx \frac{R_f}{(1 + s\tau)^2}. \quad (3.25)$$

In order to explicit the $-3dB$ frequency, we can consider:

$$R_m(f_{-3dB}) = \frac{\sqrt{2}}{2} R_f = R_f \frac{f_\tau^2}{(f_\tau^2 + f^2)}, \quad (3.26)$$

and therefore:

$$f_{-3dB} = f_\tau \left( \frac{2 - \sqrt{2}}{2} \right)^{\frac{1}{2}} \approx 0.54 \cdot f_\tau \quad (3.27)$$

This frequency is about half of the natural frequency of the system $\omega_n$ ($f_{\tau_n} = \frac{1}{2\pi\tau_n}$). If we don't ignore the zero at the numerator of equation 3.17, we have the trans-impedance:

$$R_m(s) = \frac{R_f G_0}{1 + G_0} \frac{(1 - s\tau_z)}{(1 + s\tau)^2} \quad (3.28)$$

where $\tau_z = R^* C_f / G_0$ (time constant corresponding to the TIA frequency of the zero) and $G_0 = (g_m R^* - \frac{R^*}{R_f})$ (DC gain). Equation 3.28 is at the basis of our next analysis about the timing performance of the TIA circuit.

## 4 Analysis and characteristics of the TIA response

The trans-impedance in the $s$-domain $R_m(s)$ (equation 3.28) is the TIA transfer function ($\mathcal{T}$). This needs to be convoluted with the detector current $I_D(s)$ in order to get the output voltage of the circuit. We consider here as an example the simplified condition of a 3D-trench sensor operating with charge carriers both at saturation velocities. In this case, the current has a shape that can be modeled as a simple rectangular pulse, having a width of duration $t_c$ (where $t_c$ is the charge collection time) and an amplitude $I_0$, such that the product $I_0 \cdot t_c$ equals the total charge $Q_{in}$ deposited by a particle (figure 3). This solution doesn't take into account the different drift velocities of the carries but it is still a more realistic description compared to describing the current pulse as a simple Dirac delta. The current can then be expressed in the $s$-domain as:

$$I_D(s) = I_0 \frac{1 - e^{-st_c}}{s} \quad (4.1)$$

in the time domain it can be written as the product of two Heaviside step functions:

$$I_D(t) = I_0 \theta(t) \theta(t_c - t) \quad (4.2)$$

The output voltage $V_{out}(s)$ from the circuit with $\mathcal{T} = R_m(s)$ can be written:

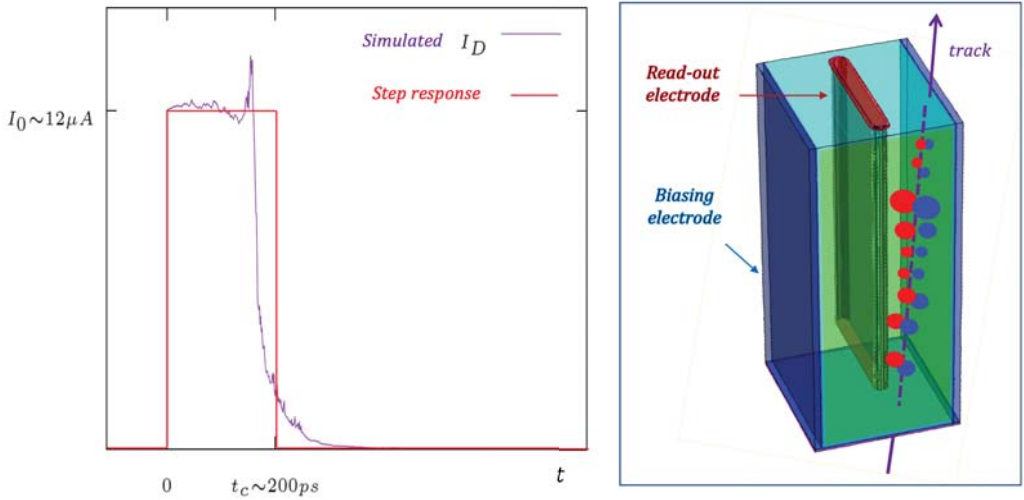

**Figure 3**. Current pulse $I_D(t)$ (left) for a 3D pixel sensor with trench geometry (right). The simulated signal is obtained by TCoDe simulation [5]. The sizes of the pixel are $55 \times 55 \times 150\ \mu m^3$.

$$V_{out}(s) = I_0 \frac{1 - e^{-st_c}}{s} \frac{R_f G_0}{1 + G_0} \frac{(1 - s\tau_z)}{(1 + s\tau)^2} \quad (4.3)$$

taking the inverse Laplace transform in the time domain we have the signal:

$$V_{out}(t) = \mathcal{L}^{-1}(t)\left\{I_0 \frac{1 - e^{-st_c}}{s} \frac{R_f G_0}{1 + G_0} \frac{(1 - s\tau_z)}{(1 + s\tau)^2}\right\} \quad (4.4)$$

The solution is:

$$V_{out}(t) = I_0 \frac{R_f G_0}{1 + G_0}\left\{\left[1 - e^{-\frac{t}{\tau}}\left(1 + \frac{t}{\tau}\left(1 + \frac{\tau_z}{\tau}\right)\right)\right] - \theta(t - t_c)\left[1 - e^{-\frac{(t-t_c)}{\tau}}\left(1 + \frac{(t - t_c)}{\tau}\left(1 + \frac{\tau_z}{\tau}\right)\right)\right]\right\} \quad (4.5)$$

We now proceed to analyse the behavior of the TIA circuit by considering separately two different operating conditions, distinguished by the size of the circuit time constant $\tau$ with respect to the charge collection time $t_c$. As already discussed above, in both cases we choose to consider the system to operate in a *critically dumped* condition ($\zeta = 1$, see equations 3.21 and 3.22).

### 4.1 Condition I: $\zeta \approx 1$ and $\tau \gg t_c$ (CS-TIA).

This condition is typical of a CSA where the value of the feedback resistor $R_f$ is maximized to have a better Signal over Noise ratio (SNR). This is an optimal configuration when the precision in the signal amplitude measurement is important at the expenses of time resolution. In any case, the use of the CSA configuration often remains a convenient compromise between overall performance and power consumption.

The bandwidth of the TIA is kept much smaller compared to the bandwidth of the current pulse and consequently the shape of the current signal is not preserved. With a given trans-conductance $g_m$ of the input transistor, the output voltage reaches quickly the maximum achievable slope that then decreases exponentially with time. When $t < t_c$, we can ignore the factor $\theta(t - t_c)$ in equation 4.2, because this term is still not contributing. We therefore get the output signal:

$$V_{out}(t)_{t<t_c} = I_0 \frac{R_f G_0}{1 + G_0} \left\{ \left[ 1 - e^{-\frac{t}{\tau}} \left( 1 + \frac{t}{\tau} \left( 1 + \frac{\tau_z}{\tau} \right) \right) \right] \right\} \tag{4.6}$$

which has derivative:

$$V'_{out}(t)_{t<t_c} = I_0 \frac{R_f G_0}{1 + G_0} \left\{ \frac{e^{-\frac{t}{\tau}}}{\tau} \left( \frac{t}{\tau} \left( 1 + \frac{\tau_z}{\tau} \right) - \frac{\tau_z}{\tau} \right) \right\} \tag{4.7}$$

This derivative equals zero at time $t_1$:

$$V'_{out}(t_1)_{t<t_c} = 0 \longrightarrow t_1 = \frac{\tau \tau_z}{\tau + \tau_z} \tag{4.8}$$

In the CSA, the zero frequency is much smaller than the one corresponding to its poles, therefore $\tau \gg \tau_z$ and $t_1 \sim \tau_z$. Substituting the $t_1$ expression into equation 4.6 we get the voltage:

$$V_{out}(t_1)_{t<t_r} \approx I_0 \frac{R_f G_0}{1 + G_0} \left( 1 - e^{-\frac{\tau_z}{\tau}} \frac{(\tau_z^2 + \tau \tau_z + \tau^2)}{\tau^2} \right) \tag{4.9}$$

This value is negative for current pulses with $I_0 > 0$. A sinking current from the sensor leads to a negative voltage at the input of the circuit and, since we have an inverting amplifier, the output voltage has a positive edge. As a consequence, the output voltage has to be negative before the total charge is collected and becomes positive only for $t > t_c$.

We now consider the second part of the equation 4.2, when $\theta(t - t_c) = 1$. The output signal expression becomes:

$$V_{out}(t)_{t>t_c} = I_0 \frac{R_f G_0}{1 + G_0} e^{-\frac{t}{\tau}} \left( \frac{t}{\tau} (e^{\frac{t_c}{\tau}} - 1)(\frac{\tau_z + \tau}{\tau}) + \frac{e^{\frac{t_c}{\tau}} (\tau^2 - t_c(\tau - \tau_z))}{\tau^2} - 1 \right) \tag{4.10}$$

We can define:

$$A = I_0 \frac{R_f G_0}{1 + G_0} \qquad B = (e^{\frac{t_c}{\tau}} - 1)\left( \frac{\tau_z + \tau}{\tau} \right) \qquad C = \frac{e^{\frac{t_c}{\tau}} (\tau^2 - t_c(\tau - \tau_z))}{\tau^2} - 1$$

and therefore equation 4.10 can be written as:

$$V_{out}(t)_{t>t_c} = A e^{\frac{-t}{\tau}} \left( B \frac{t}{\tau} + C \right) \tag{4.11}$$

Taking the derivative of this expression, we find the peaking time $T_{peak}$:

$$T_{peak} = \frac{B - C}{B} \tau \tag{4.12}$$

$$T_{peak} = \frac{e^{\frac{t_c}{\tau}}(\tau_z(\tau - t_c) + \tau t_c) - \tau\tau_z}{(\tau_z + \tau)(e^{\frac{t_c}{\tau}} - 1)} \quad (4.13)$$

since $\tau \gg t_c$ we can take the limit for $t_c \longrightarrow 0$, obtaining:

$$T_{peak} = \frac{\tau^2}{(\tau_z + \tau)} \approx \tau \quad (4.14)$$

The output signal $V_{out}$ is plotted in figure 4. During charge collection, the signal is negative. At $t > t_c$ the signal becomes positive with positive derivative, reaching a maximum at $T_{peak} \approx \tau$. We can anticipate here that this condition does not appear as the best possible one when the speed of the sensor is to be fully exploited. We will come back extensively on this point in section 6. We now analyze in further detail some relevant characteristics of the calculated circuit responses given in equations 4.11 and 4.10.

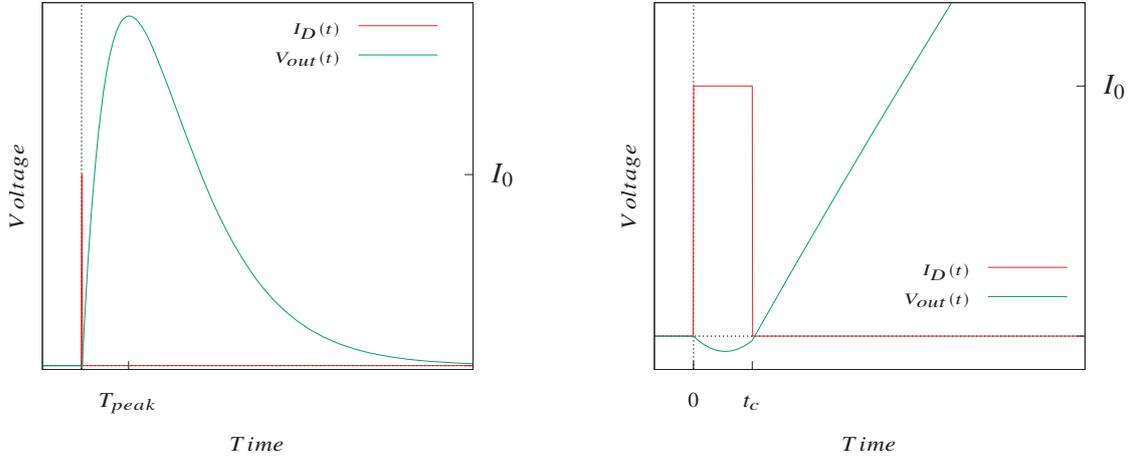

**Figure 4**. Calculated output voltage $V_{out}$ in the CS-TIA configuration. Right: Due to the high peaking time ($T_{peak} \approx \tau$) with respect to average collection time, the current signal can be approximated by a Delta function. Right: Detail of the under-shoot during $t < t_c$.

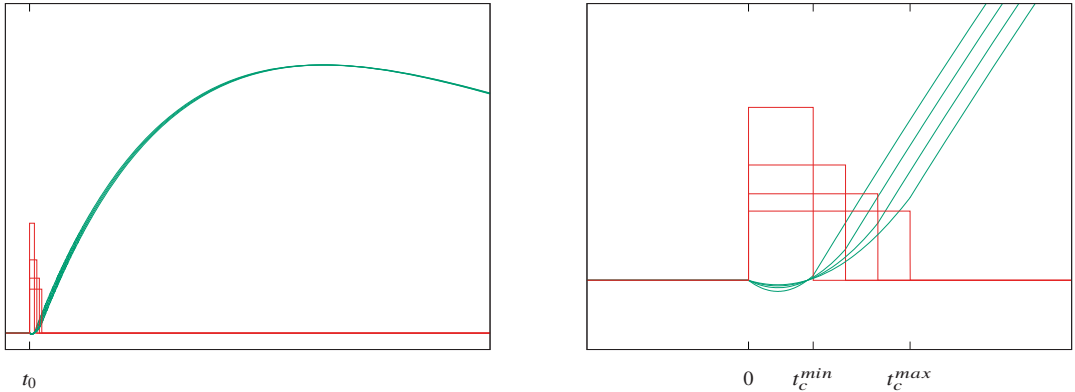

**Figure 5**. Output voltage $V_{out}$ in the CS-TIA configuration for current pulses with different duration $t_c$ and same charge $Q_{in} = I_0 \cdot t_c$ (left). Detail of the under-shoot and slope of the signal for different $t_c$

### 4.1.1 Slope of $V_{out}$ in CS-TIA

If the current pulse duration is much shorter compared to the time constant ($\tau \gg t_c$), we have that the slope of the signal after the induction is almost independent from $t_c$. As shown in figure 5, different charge collection times lead to a delay of the signals but the initial slope is about the same. The maximum slope for every current is reached at time $t_c$ and then decreases exponentially. We can consider equation 4.10 to calculate the signal slope for $t > t_c$. The signal derivative is:

$$V'_{out}(t)_{t>t_c} = -\frac{V_{out}(t)}{\tau} + \frac{A \cdot B e^{-\frac{t}{\tau}}}{\tau} \tag{4.15}$$

setting $t = t_c$ and using $e^{\frac{t_c}{\tau}} \approx 1$ we find:

$$V'_{out}(t_c) \approx \frac{A \cdot B}{\tau}(1 - \frac{t_c}{\tau} - \frac{C}{B}) \approx \frac{A \cdot B}{\tau}(1 - \frac{t_c}{\tau} + \frac{\tau_z}{\tau_z + \tau}) \tag{4.16}$$

the terms $\frac{t_c}{\tau}$ and $\frac{\tau_z}{\tau_z+\tau}$ are small and with opposite sign, therefore we can write:

$$V'_{out}(t_c) \approx \frac{A \cdot B}{\tau} = \frac{I_0 R_f G_0}{1+G_0}(e^{\frac{t_c}{\tau}} - 1)\left(\frac{\tau_z + \tau}{\tau^2}\right) \approx \frac{I_0 R_f G_0}{1+G_0}\frac{t_c}{\tau}\left(\frac{\tau_z + \tau}{\tau^2}\right) \tag{4.17}$$

$$V'_{out}(t_c) \approx \frac{I_0 R_f G_0}{1+G_0}\frac{t_c}{\tau}\left(\frac{\tau_z + \tau}{\tau^2}\right) \approx \frac{G_0}{1+G_0}Q_{in}R_f\left(\frac{1}{\tau \cdot T_{peak}}\right) \tag{4.18}$$

Being $T_{peak} \approx \tau$ (equation 4.14) and considering that $\frac{G_0}{1+G_0} \approx 1$ we can write the slope of the signal as:

$$V'_{out}(t_c) \approx \frac{Q_{in}R_f}{\tau^2} \tag{4.19}$$

using equation 3.23 and the expression of $\omega_n$ found in equation 3.20:

$$\frac{dV}{dt} \approx \frac{Q_{in} \cdot g_m}{\xi} \tag{4.20}$$

$$\frac{dV}{dt} \approx \frac{Q_{in} \cdot g_m}{(C_L C_{in} + C_L C_f + C_{in} C_f)} \tag{4.21}$$

The slope of the signal can be estimated just considering the input charge $Q_{in}$, the trans-conductance $g_m$ of the first amplifier stage and the circuit capacitances through the quantity $\xi$.

### 4.1.2 Output Voltage Noise $\sigma_V$ in CS-TIA

We can identify three noise sources in the TIA (figure 1, right). Two sources are due to the resistances $R_f$ and $R_D$, and a third source is due to the MOS transistor and more precisely depends on its trans-conductance $g_m$ and $\gamma$ factor ($\gamma \sim \frac{2}{3}$ but becomes higher for deep sub-micron technologies). The load can be provided by another MOS transistor and in that case the noise will also depend on the trans-conductance of the load element. Figure 6 shows the circuit with its identified noise sources [7].

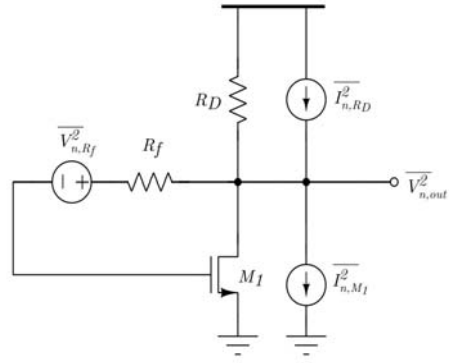

**Figure 6**. noise sources of the Feedback TIA

The calculation of the noise can be found in Appendix A, here we report the results:

$$\sigma_{v,M_1} = \sqrt{\frac{k_B T \gamma}{2}} \frac{C_f + C_{in}}{\xi^{\frac{3}{4}}} g_m^{\frac{1}{4}} R_f^{\frac{1}{4}} \tag{4.22}$$

$$\sigma_{v,R_D} = \sqrt{\frac{k_B T}{2 R_D}} \frac{C_f + C_{in}}{g_m^{\frac{1}{4}} \xi^{\frac{3}{4}}} R_f^{\frac{1}{4}} \tag{4.23}$$

$$\sigma_{v,R_f} = \sqrt{\frac{k_B T}{2}} \frac{g_m^{\frac{1}{4}}}{R_f^{\frac{3}{4}} \xi^{\frac{1}{4}}} \tag{4.24}$$

where $\xi = C_f C_{in} + C_f C_L + C_L C_{in}$.

The transistor noise is proportional to the trans-conductance $g_m$. It is known that increasing $g_m$ by an increase of the bias current leads to a smaller output resistance of the transistor that decreases the voltage gain and the output noise. In this case, the noise value is considered keeping the circuit in a condition where the damping factor $\zeta$ stays equals unity so that we are in a critically damped system. Changing one parameter (for example $R_f$) leads to a change in both the natural frequency and damping factor. The effect of this on noise must be carefully considered as it changes the power spectral density of the noise sources (figure 7).

The total noise is given adding in quadrature the three noise sources:

$$\sigma_{v,TOT} = \sqrt{\sigma_{v,M_1}^2 + \sigma_{v,R_f}^2 + \sigma_{v,R_D}^2} \tag{4.25}$$

The voltage peak can be calculated using the $T_{peak}$ expression given in equation 4.13 and is given by:

$$V_{out}^{peak} \approx V_{out}(\tau)_{t>t_c} = A e^{-1} \left( B + C \right) \tag{4.26}$$

replacing $A, B, C$ with the values defined in equation 4.1 and considering $e^{\frac{t_c}{\tau}} \approx (1 - \frac{t_c}{\tau})$ we get:

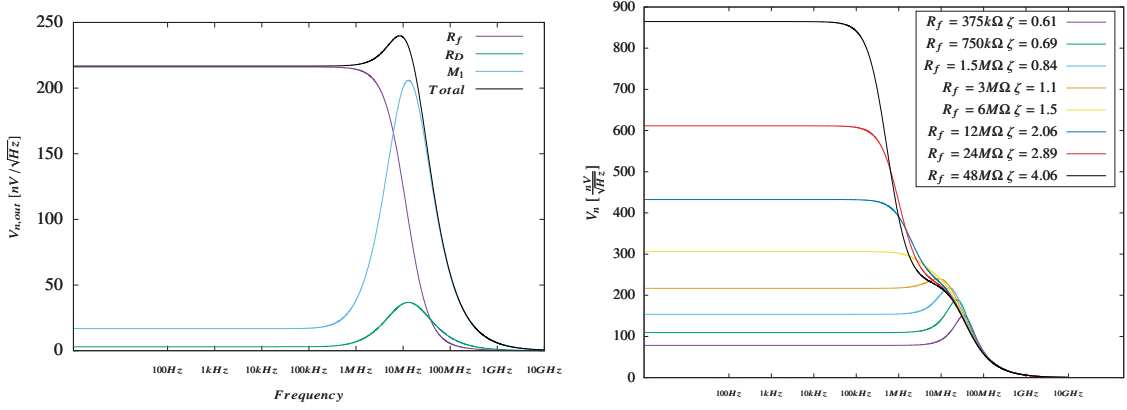

**Figure 7**. Left: example of Noise Power Spectral Density contributions of $R_f$, $R_D$ and transistor $M_1$. Right: output noise for different damping factor $\zeta$ obtained by changing the value of $R_f$.

$$V_{out}^{peak} \approx \frac{G_0}{1+G_0} \frac{Q_{in} R_f}{e} \left( \frac{\tau(\tau - t_c) - t_c \tau_z}{\tau^3} \right) \quad (4.27)$$

$$V_{out}^{peak} \approx \frac{Q_{in} R_f}{e\tau} \quad (4.28)$$

As an example, considering only the noise contribution of $M_1$ the SNR can be written:

$$SNR_{out,M_1} \approx \sqrt{\frac{2}{k_B T \gamma}} \frac{Q_{in}}{e(C_{in}+C_f)} (g_m R_f \xi)^{\frac{1}{4}} \quad (4.29)$$

The expressions obtained in this section will be the main ingredients for our discussion on timing performance of the CSA which we will cover below.

### 4.1.3 Front-end Jitter in CS-TIA

Starting from equations 4.19 and 5.3, the CSA jitter can be written:

$$\sigma_j = \frac{\sigma_V \tau^2}{Q_{in} R_f} \quad (4.30)$$

Using equation 4.21, we can make clear the trans-conductance $g_m$ and the circuit capacitances through the quantity $\xi = C_L C_{in} + C_L C_f + C_f C_{in}$:

$$\sigma_j = \frac{\sigma_V \xi}{Q_{in} g_m} \quad (4.31)$$

Considering the noise calculated in subsection 4.1.2, we can calculate the contribution to the time resolution given by the single noise sources:

$$\sigma_{j,M_1} = \sqrt{\frac{k_B T \gamma}{2}} \frac{C_f + C_{in}}{Q_{in} \cdot g_m^{\frac{3}{4}}} (\xi R_f)^{\frac{1}{4}} \quad (4.32)$$

– 13 –

$$\sigma_{j,R_f} = \sqrt{\frac{k_B T}{2}} \frac{\xi^{\frac{3}{4}}}{Q_{in} R_f^{\frac{1}{4}} g_m^{\frac{3}{4}}} \tag{4.33}$$

$$\sigma_{j,R_D} = \sqrt{\frac{k_B T}{2R_D}} \frac{C_f + C_{in}}{Q_{in} \cdot g_m^{\frac{5}{4}}} \left(\frac{R_f}{\xi}\right)^{\frac{1}{4}} \tag{4.34}$$

summing in quadrature to find the total jitter:

$$\sigma_{j,TOT} = \sqrt{\sigma_{j,M_1}^2 + \sigma_{j,R_f}^2 + \sigma_{j,R_D}^2} \tag{4.35}$$

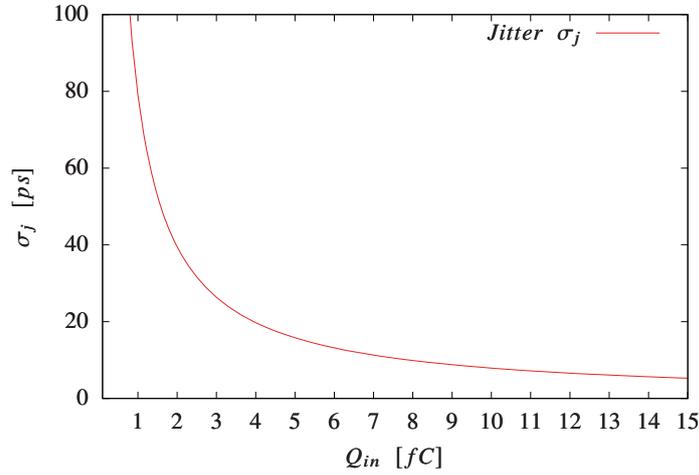

**Figure 8**. Jitter estimate for different $Q_{in}$ using equation 4.31.

Equation 4.35 estimates the minimum jitter that can be reached in this topology. Since the derivative is taken at time $t = t_c$, we are considering the maximum slope when the signal is still close to zero. in the real case, we need to set a threshold with a value higher than the noise. Therefore, the jitter depends on the value chosen for the threshold ad since the slope decreases exponentially, a higher threshold means higher jitter. Equation 4.31 is still useful to have a rough evaluation of the performance achievable with a FB-TIA configuration such that $\tau \gg t_c$. Considering as an example a circuit with having $\tau = 12ns$, feedback resistance $R_f = 3M\Omega$, $SNR = 100$, with an input charge $Q_{in} = 2fC$ the achievable jitter is in the order of 40 $ps$ (figure 8).

### 4.1.4 Example of CS-TIA circuit optimization

In the present subsection, we use the jitter expressions above to suitably size the value of $g_m$ for minimum jitter, while still respecting the stability conditions of the amplifying stage.

Supposing the most significant noise source is given by the transistor $M_1$, then we can find the optimum jitter solving the following system:

$$Jitter_{opt} \begin{cases} \sigma_{j,M_1} < \sigma_{j,Target} \\ \zeta = 1 \end{cases} \quad (4.36)$$

$$Jitter_{opt} \begin{cases} \sqrt{\frac{k_B T \gamma}{2}} \frac{C_f + C_{in}}{Q_{in} \cdot g_m^{\frac{3}{4}}} (\xi R_f)^{\frac{1}{4}} < \sigma_{j,Target} \\ \frac{R_f (C_{in} + C_f(1+G_0)) + R^* C_L}{\sqrt{(1+G_0) R^* R_f \xi}} = 2 \end{cases} \quad (4.37)$$

the first condition ask that the jitter is lower then the target value that is needed, while the second equation is to make sure that the system stay stable with a critically damped behavior. Introducing $g_m$ in the second equation we get:

$$Jitter_{opt} \begin{cases} \sqrt{\frac{k_B T \gamma}{2}} \frac{C_f + C_{in}}{Q_{in} \cdot g_m^{\frac{3}{4}}} (\xi R_f)^{\frac{1}{4}} < \sigma_{j,Target} \\ \frac{R_f (C_f(1+g_m R_D)) + R_D C_L + (R_f + R_D) C_{in}}{\sqrt{(1+g_m R_D) R_D R_f \xi}} = 2 \end{cases} \quad (4.38)$$

Supposing we want to decrease the jitter doubling $g_m$, the first equation tells us that the jitter decreases by a factor of $2^{\frac{3}{4}} \approx 1.68$ while the second equation implies that the damping factor increases by a factor of about $\sqrt{2} \approx 1.414$ ($\zeta \propto \sqrt{g_m}$). In this case we would have to decrease $R_f$ by the same factor to have still a critically damped system (since $\zeta \propto \sqrt{R_f}$). If we double $g_m$ and halve $R_f$ the jitter becomes:

$$\sigma_{j,M_1,new} \approx \frac{1}{2^{1/4} \cdot 2^{3/4}} \sigma_{j,M_1,old} = \frac{\sigma_{j,M_1,old}}{2} \quad (4.39)$$

Assuming that the values of the other components $C_D, R_D, C_f, C_L$ don't change, we would end up with half the jitter due to $M_1$. Changing the biasing changes the output resistance of the transistor ($r_0$) that is in parallel with $R_D$ ($r_0 \propto \frac{1}{\lambda I_{bias}}$, where $\lambda$ is the channel lenght modulation coefficient). This has to be taken into account in the specific case. Following this method, it is possible for example to make explicit the design parameters $W, L$ and optimize the size of the input transistor. Considering the other two jitter contributions we find:

$$\sigma_{j,R_D,new} \approx \frac{1}{2^{1/4} \cdot 2^{5/4}} \sigma_{j,R_D,old} = \frac{\sigma_{j,R_D,old}}{2^{\frac{3}{2}}} \quad (4.40)$$

the jitter due to $R_f$ becomes:

$$\sigma_{j,R_f,new} \approx \frac{2^{1/4}}{2^{3/4}} \sigma_{j,R_f old} = \frac{\sigma_{j,R_f,old}}{\sqrt{2}} \quad (4.41)$$

doubling the $g_m$ and halving the feedback resistance decreases all the three jitter contributions that we are considering.

## 4.2 Condition II: $\zeta \approx 1$ and $\tau \approx t_c$ (Fast-TIA).

A possible solution for realizing a FB-TIA having a time constant of the same order of the charge collection time $t_c$ is using the same self-biased scheme of figure 1, but implemented with a high-bandwidth Si-Ge bipolar transistor stage, as illustrated in figure 9. It is thus possible to take advantage of the benefits of the Si-Ge devices that allow small input capacitances and high-frequency transitions of the order of 100 GHz also for discrete-component circuit solutions.

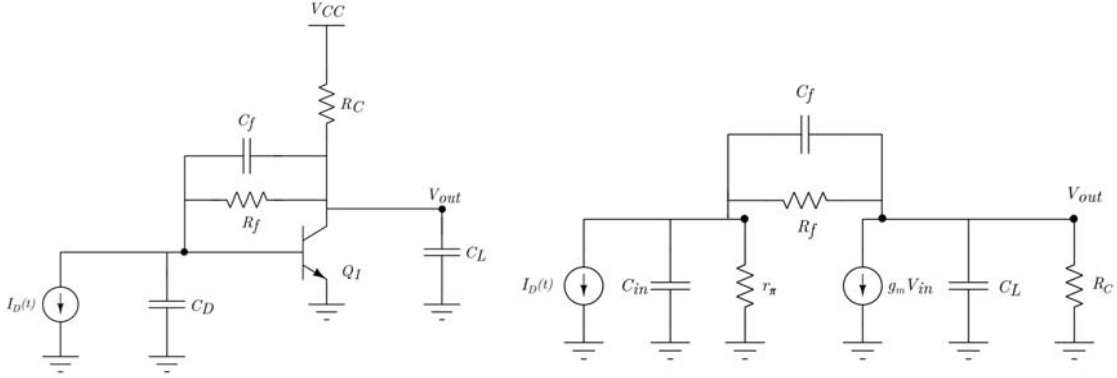

**Figure 9**. Schematic of the FB-TIA with bipolar transistor NPN (left) and corresponding Small signal model (right).

The difference with the MOS solution is that now a bias current $I_b$ flows through the feedback resistance $R_f$, giving a dynamic input resistance $r_\pi$, defined as:

$$r_\pi = \frac{V_T}{I_b} \quad (4.42)$$

where $V_T = \frac{k_B T}{e} \approx 26mV$ at 300 K. The circuit small signal model (figure 9, right) can be used to solve the circuit in detail. We calculate the input voltage $V_{in}(s)$, the natural frequency $\omega_n$ and the damping factor $\zeta$:

$$V_{in}(s) = -I_D \frac{r_\pi(R_f + R_C + sR_f R_C(C_f + C_L))}{(R_f R_C r_\pi \xi)(s^2 + 2\zeta\omega_n s + \omega_n^2)} \quad (4.43)$$

$$\omega_n = \sqrt{\frac{r_\pi(1 + g_m R_C) + R_C + R_f}{R_f R_C r_\pi \xi}} \quad (4.44)$$

$$\zeta = \frac{1}{2}\left\{\frac{r_\pi C_{in}(R_f + R_C) + R_f C_f(r_\pi(1 + g_m R_C) + R_C) + R_C C_L(R_f + r_\pi)}{\sqrt{(r_\pi(1 + g_m R_C) + R_C + R_f)R_f R_C r_\pi \xi}}\right\} \quad (4.45)$$

The zero found in the voltage gain (equation 3.10) can be neglected due to the very high trans-conductance $g_m$ of our Si-Ge BJT, which takes the frequency corresponding to the zero time constant $\tau_z$ to extremely high values. The voltage gain is then given by:

$$G_v(s) = -\frac{(g_m R_f - 1)R_C}{R_f + R_C + sR_C R_f(C_f + C_L)}. \quad (4.46)$$

Also in this case, we can simplify the transfer function considering a critically dumped behavior ($\zeta \approx 1$). The output voltage becomes:

$$V_{out}(s) = I_D \frac{r_\pi R_C (g_m R_f - 1)}{(R_f + R_C + r_\pi(1 + g_m R_C))(1 + s\tau)^2}. \tag{4.47}$$

Defining the trans-impedance:

$$R_{m_0} = \frac{r_\pi g_m R_C R_f - r_\pi}{(R_f + R_C + r_\pi(1 + g_m R_C))}, \tag{4.48}$$

that can be approximated using $R^* = R_f || R_C$ and $R_{in} = \frac{r_\pi(R_f + R_C)}{(R_f + R_C + r_\pi(1 + g_m R_C))}$:

$$R_{m_0} \approx g_m R^* R_{in} \tag{4.49}$$

we finally obtain the trans-impedance of the circuit:

$$R_m(s) = \frac{R_{m_0}}{(1 + s\tau)^2} \tag{4.50}$$

The output voltage is given by the convolution with the current pulse defined in equation 4.2, and reads:

$$V_{out}(t) = \mathcal{L}^{-1}(t) \left\{ I_0 \frac{1 - e^{-s t_c}}{s} \frac{R_{m_0}}{(1 + s\tau)^2} \right\} \tag{4.51}$$

for $\zeta = 1$ e $t_c = \tau$ we obtain the voltage output given by:

$$V_{out}(t) = I_0 R_{m_0} \left\{ \left[ 1 - e^{-\frac{t}{\tau}} \left(1 + \frac{t}{\tau}\right) \right] - \theta(t - t_c) \left[ 1 - e^{-\frac{(t - t_c)}{\tau}} \left(1 + \frac{(t - t_c)}{\tau}\right) \right] \right\} \tag{4.52}$$

### 4.2.1 Slope of $V_{out}$ in Fast-TIA

When the time constant of the circuit $\tau$ is of the same order of the charge collection time $t_c$, the front-end output signal can reach the maximum slope before the charge is completely collected at the sensor electrodes. We can demonstrate this taking the derivative of the first part of the solution 4.52 which reads:

$$V'_{out}(t)_{t < t_c} = \frac{I_0 R_{m_0}}{\tau^2} e^{-\frac{t}{\tau}} t \tag{4.53}$$

Taking the second derivative we get:

$$V''_{out}(t)_{t < t_c} = \frac{I_0 R_{m_0}}{\tau^2} e^{-\frac{t}{\tau}} \left(1 - \frac{t}{\tau}\right) \tag{4.54}$$

the maximum slope is reached at time $t = \tau$ and since we are implying $t_c = \tau$ we have the slope:

$$\left(\frac{dV}{dt}\right)_{max} = \frac{Q_{in} R_{m_0}}{e\tau^2}. \tag{4.55}$$

Using the definitions of $\tau$ and $R_{m_0}$ we find:

$$\left(\frac{dV}{dt}\right)_{max} = \frac{Q_{in}(g_m - \frac{1}{R_f R_C})}{\xi e}. \tag{4.56}$$

Usually $\frac{1}{R_f R_C} \ll g_m$ and we have a solution similar to the MOS transistor case with $\tau \gg t_c$, with the difference of the factor $e \sim 2.71$ at the denominator:

$$\left(\frac{dV}{dt}\right)_{max} = \frac{Q_{in} g_m}{\xi e} \tag{4.57}$$

To find the peaking time $T_{peak}$ and voltage peak $V_{peak}$ we consider the solution for $t > t_c$:

$$T_{peak} = \tau \frac{e}{e-1} \approx 1.582 \cdot \tau \tag{4.58}$$

$$V_{peak} = I_0 R_{m_0} e^{-\left(\frac{e}{e-1}\right)}(e-1) \tag{4.59}$$

$$V_{peak} \approx 0.353 \cdot I_0 R_{m_0} \tag{4.60}$$

The voltage at time $t = \tau$, that is the maximum slope voltage, is:

$$V_{\substack{Max \\ Slope}} = I_0 R_{m_0}\left(\frac{e-2}{e}\right) \approx 0.264 \cdot I_0 R_{m_0} \tag{4.61}$$

taking the ratio between $V_{\substack{Max \\ Slope}}$ and $V_{peak}$ we find:

$$V_{\substack{Max \\ Slope}} \approx 0.748 \cdot V_{peak} \tag{4.62}$$

In our condition of $t_c = \tau$, the maximum slope is reached at about 75% of the peaking value (figure 10). This means that for an amplifier with short time constant the threshold has to be set to an higher value compared to the CSA solution to minimize the electronic jitter (see also section 6).

### 4.2.2 Noise in Fast-TIA

The noise introduced by a bipolar transistor is due to two correlated shot noise sources which cause fluctuations in the bias currents $I_b$ and $I_c$. The Power Spectral Densities (PSD) for this two sources are:

$$I_{n,b}^2 = \frac{2k_B T}{r_\pi}$$

$$I_{n,c}^2 = 2k_B T g_m \tag{4.63}$$

the other two noise sources are given by the resistor $R_f$ and $R_D$. For the calculation of the single noise contributions see appendix B, here we report the results:

$$\sigma_{v,b}^2 = \frac{k_B T}{4 r_\pi \tau}(g_m R^* R_{in})^2 \tag{4.64}$$

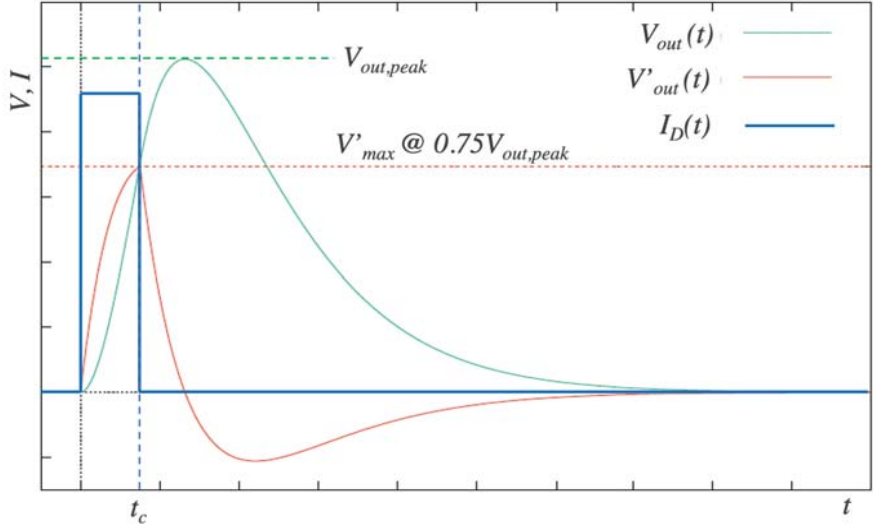

**Figure 10**. Output signal and its derivative for a FB-TIA in Condition II ($\tau \approx t_c$ and $\zeta \approx 1$)

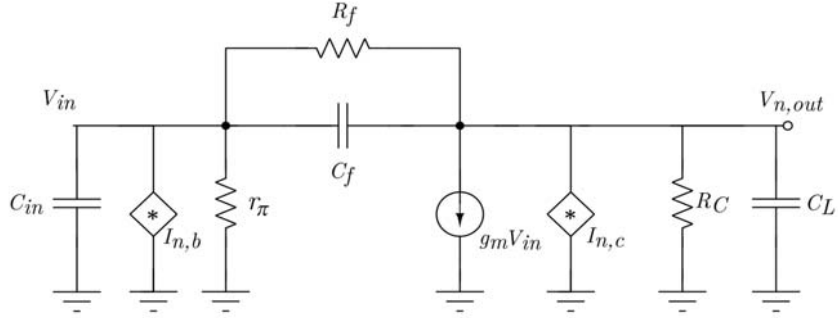

**Figure 11**. Small signal model for the noise sources of a self-biased circuit implemented with BJT

$$\sigma_{v,c}^2 = \frac{k_B T g_m}{4\tau}\left(R_{out}^2 + \frac{R^* R_{in}(C_{in} + C_f)^2}{\xi}\right) \qquad (4.65)$$

$$\sigma_{v,R_C}^2 \approx \frac{k_B T}{2\tau R_C}\left(R_{out}^2 + R^* R_{in}\frac{C_{in}}{C_L}\right) \qquad (4.66)$$

$$\sigma_{v,R_f}^2 = \frac{k_B T R_f}{(1 + \frac{R_f + R_C}{r_\pi g_m R_C})^2}\frac{1}{2\tau} \qquad (4.67)$$

As an example the signal to noise ratio ($SNR$) considering only the source $I_{n,b}^2$ and the expression for $V_{peak}$ (eq. 4.60) is given by:

$$SNR \approx 0.7 \cdot I_0 \sqrt{\frac{r_\pi \tau}{k_B T}} \qquad (4.68)$$

### 4.2.3 Front-end Jitter in Fast-TIA

In the case of $\tau \sim t_c$ and large bandwidth, usually the most significant noise source is given by the fluctuations of the bias current $I_b$. The jitter given by this source can be written:

$$\sigma_{j,b} = \sqrt{\frac{k_B T}{4 r_\pi}} \frac{e}{Q_{in}} \tau^{\frac{3}{2}} \qquad (4.69)$$

the dependence on $g_m$, (and therefore on power consumption), is implicit since the time constant $\tau$ depends on $g_m$. Also in this case we have that the jitter is proportional to $g_m^{-\frac{3}{4}}$. however, for BJT-TIA implementation, the relationship with power consumption is more complicated than in the MOS case. For a bipolar transistor, higher trans-conductance $g_m$ means higher bias current $I_b$ that causes the dynamic resistance $r_\pi$ to decrease (eq. 4.42). This means that the time constant changes together with the damping factor $\zeta$, thus changing the damping behavior of the circuit. Using Eq. 4.69, we can obtain an estimate of the achievable jitter when: $\tau \sim t_c$, $\zeta \sim 1$ and the threshold is set to $V_{\frac{Max}{Slope}}$.

### 4.2.4 Example of Fast-TIA circuit

A clearer explanation of what obtained in the previous sections can be given using a circuit example with given specifications. Referring to figure 9, let's consider a circuit with the following component values:

$$R_f = 4k\Omega \quad R_C = 35\Omega \quad g_m = 0.78S \quad r_\pi = 300\Omega$$

$$C_D = 1pF \quad C_{be} = 0.5pF \quad C_f = 60fF \quad C_L = 6pF$$

$$C_{in} = 1.5pF \quad R^* = 34.69\Omega \quad R_{in} = 96.64\Omega \quad R_{out} = 12.01\Omega$$

with these values we have the following natural frequency and damping factor:

$$\omega_n = 5.61 \cdot 10^9 rad/s \quad \tau = 178ps \quad \zeta = 1.09$$

If the current pulse has $t_c = 150\ ps$, $Q_{in} = 2fC$, $I_0 = \frac{Q_{in}}{t_c} = 13.33\mu A$, while the trans-impedance is $R_{m_0} = 2.61 k\Omega$. We have then the signal:

$$V_{out}(t) = \begin{cases} I_0 R_{m_0}(1 - e^{-\frac{t}{\tau}}(1 + \frac{t}{\tau})) & \text{if } t < t_c \\ I_0 R_{m_0}(e^{-\frac{t-t_c}{\tau}}(1 + \frac{t-t_c}{\tau}) - e^{-\frac{t}{\tau}}(1 + \frac{t}{\tau})) & \text{if } t > t_c \end{cases} \qquad (4.70)$$

Figure 12 (left) shows the output signal. Since we have $t_c$ slightly smaller than $\tau$, the maximum slope is reached before the time $t = \tau$ and exactly at time $t = t_c$. The voltage $V_{peak}$ equation 4.60 is smaller compared to the condition $t_c = \tau$:

$$V_{peak} \approx 10.24 mV \quad < \quad 0.353 \cdot I_0 R_{m_0} = 12.28 mV \qquad (4.71)$$

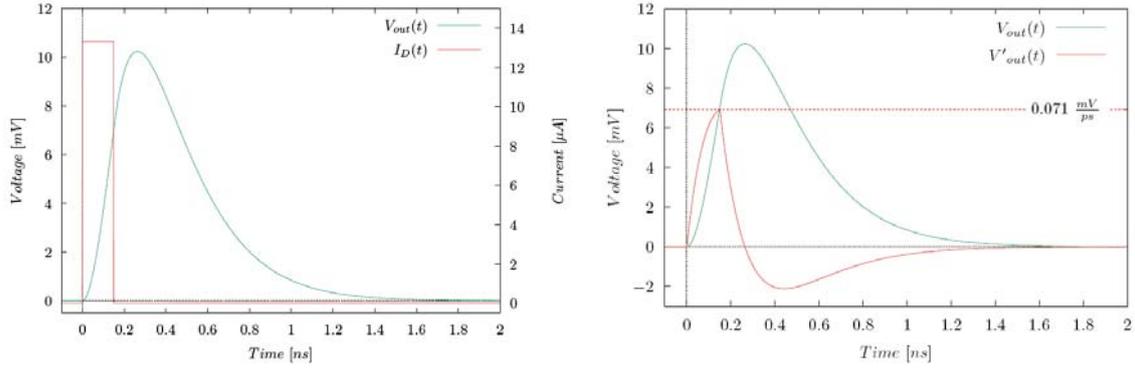

**Figure 12**. Output voltage $V_{out}(t)$ with current pulse $I_D(t)$ (left), output voltage $V_{out}(t)$ and output voltage derivative $V'_{out}(t)$ (right).

for the peaking time $T_{peak}$ we get:

$$T_{peak} = 262\,ps \quad < \quad 1.58 \cdot \tau = 281\,ps \tag{4.72}$$

for the slope of the signal we have that, since the pulse is shorter than the time constant $\tau$, the derivative has a higher value:

$$V'_{out}(t_c) = \frac{I_0 R_{m_0}}{\tau^2} t_c \cdot e^{-\frac{t_c}{\tau}} \tag{4.73}$$

$$\left(\frac{dV}{dt}\right)_{max} = \frac{Q_{in} R_{m_0}}{\tau^2} e^{-\frac{t_c}{\tau}} \approx 0.071 \frac{mV}{ps} \quad > \quad \frac{Q_{in} R_{m_0}}{\tau^2 e} \approx 0.061 \frac{mV}{ps} \tag{4.74}$$

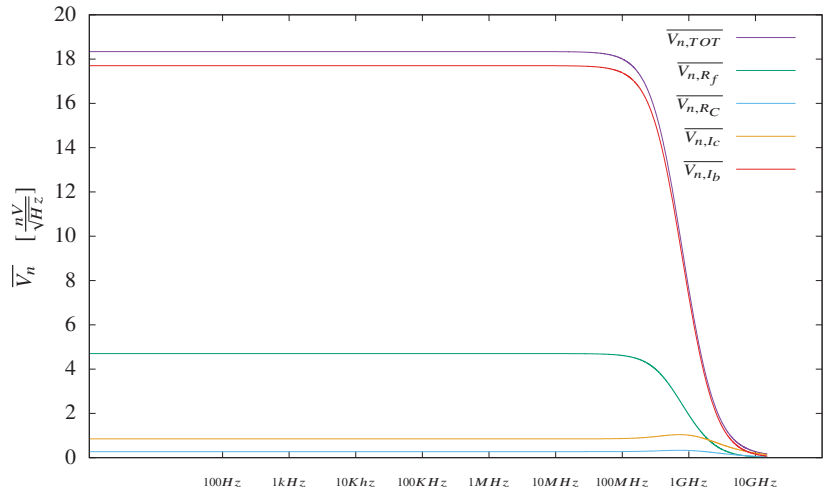

**Figure 13**. Power Spectral Density of the output noise $\sigma_{V_{out}}$

Figure 12 (right) shows the output signal together with the derivative $V'_{out}$. The derivative has a minimum corresponding to the condition of maximum descent. The noise of this circuit has the PSD shown in figure 13. The noise contributions can be specified as:

$$\sigma_{v,b} = 449.9\mu V$$

$$\sigma_{v,R_f} = 119.5\mu V$$

$$\sigma_{v,c} = 57.8\mu V$$

$$\sigma_{v,R_C} = 18.5\mu V$$

The total noise is:

$$\sigma_{v,TOT} = 469.5\mu V$$

The dominant term is $\sigma_{v,b}$ as can be seen from 13.

Using the calculated noise and the slope of the signal we can estimate the jitter of the circuit as:

$$\sigma_j = \frac{\sigma_{v,TOT}}{\frac{dV}{dt}}$$

$$\sigma_j = 6.61 ps \tag{4.75}$$

The closed loop gain of the circuit is about $28.6dB$ with a frequency cut $f_{\tau_L} = \frac{1}{2\pi R^*(C_L+C_f)} = 757MHz$. The trans-impedance $R_m$ (equation 4.50) is shown in figure 14 and has a value of $R_{m_0} = 2.61k\Omega = 68.33dB\Omega$, while the frequency cut is given by equation 3.27 and reads $f_{-3dB} \approx 483Mhz$.

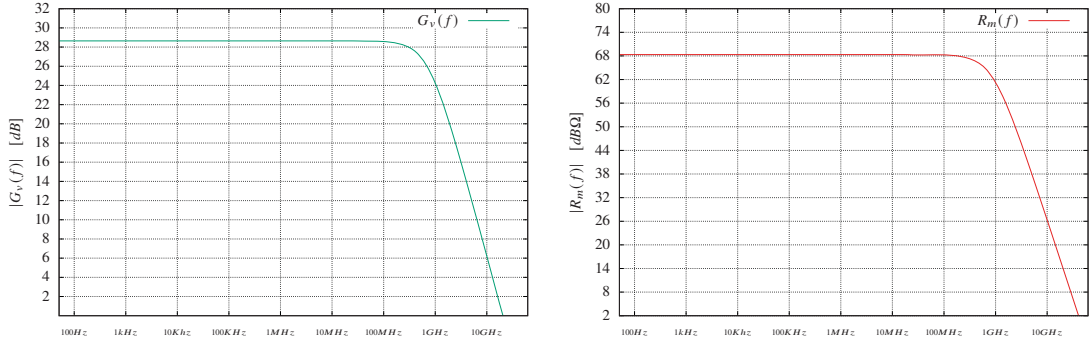

**Figure 14**. Voltage gain $G_v(f)$ (left), and trans-impedance $R_m(f)$ (right)

The transfer function $R_m(s)$ has to be convoluted with the current pulse $I_D(s)$ that in the frequency domain has the spectrum of the $Sinc(f)$ function (figure 15).

$$I_D(f) = I_0 \frac{Sin(\pi t_c f)}{\pi t_c f} \tag{4.76}$$

The trans-impedance $R_m(s)$ acts as a filter by amplifying some frequencies while suppressing others. The spectrum of the signal reach the first zero at frequency $f = \frac{1}{t_c}$ that for a pulse with $t_c = 150\ ps$ means $f \sim 6.6\ GHz$. To find the optimal amplifier time constant needed to process the significant (fast) part of the input current pulse, we can consider the $-3dB$ Bandwidth of the signal defined as:

$$\frac{|Sin(\pi t_c f)|}{|\pi t_c f|} = \frac{\sqrt{2}}{2} \tag{4.77}$$

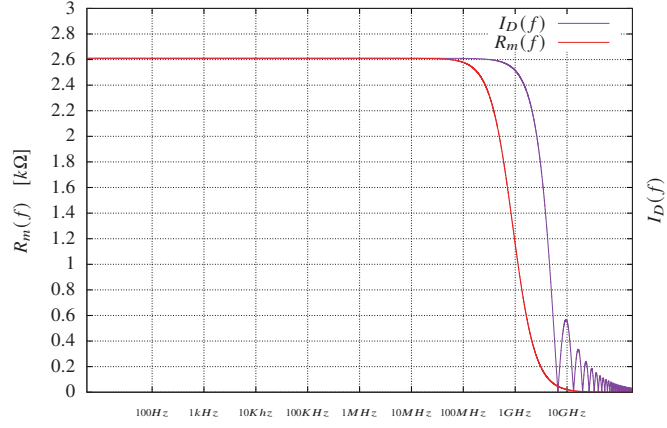

**Figure 15**. trans-impedance $R_m(f)$ and Fourier transform of $I_D(s)$

approximating $|Sin(\pi t_c f)|$ at the second order we find:

$$f_{I_D,-3dB} = \left(\frac{3(2-\sqrt{2})}{2}\right)^{\frac{1}{2}} \frac{1}{\pi t_c} \tag{4.78}$$

$$f_{I_D,-3dB} \approx \frac{0.3}{t_c} \tag{4.79}$$

for $t_c = 150~ps$ we have $f_{I_D,-3dB} \sim 2~GHz$. Since the frequency cut of the trans-impedance $R_m(s)$ is given by equation 3.27 and is about half the one defined with the natural frequency of the circuit, we have that the time constant of the circuits needs to be at least:

$$\left(\frac{3(2-\sqrt{2})}{2}\right)^{\frac{1}{2}} \frac{1}{\pi t_c} = \left(\frac{(2-\sqrt{2})}{2}\right)^{\frac{1}{2}} f_\tau \tag{4.80}$$

$$\tau = \frac{t_c}{2\sqrt{3}} \tag{4.81}$$

This means that to process the most significant part of the signal, containing its fast information, the time constant $\tau$ should be about a factor 3.4 smaller than the charge collection time $t_c$ of the sensor.

## 5 Time dispersion for sensor and front-end

In the present section we analyze the performance in time resolution of the two configuration described above, that is the CSA-like configuration ($\tau \gg t_c$) and the Fast configuration ($\tau \approx t_c$). We are specifically interested in the relationship between the front-end timing characteristics and the native time dispersion or *speed* of the sensor. This is of particular importance in case of very fast sensors, whose time distributions can have relatively small dispersion (standard deviations in the range of tens of ps).

### 5.1 Sensor contribution to time jitter

We can start considering an ideal 3D geometry with flat parallel faces (figure 16). Such specific choice is motivated by the high intrinsic speed of this kind of sensors [4]. In this case, charge collection time $t_c$ depends on the hit position of the impinging particle. For tracks closer to the electrode at the higher potential, the sensor will collect electrons very quickly while holes would induce for a longer time since they have to travel a longer distance and they move slower. The same argument can be used in the opposite case with tracks close to the electrode at lower potential with short current from holes and a longer pulse for electrons.

When we have electric fields strong enough for both charge carriers to reach the respective saturation velocities $v_e$ and $v_h$, the electrons and holes speeds becomes similar (figure 17). We will have then a minimum $t_c^{min}$ and a maximum $t_c^{max}$ for the charge collection time $t_c$. Assuming that the distance between the electrodes is $d = 20~\mu m$ we obtain:

$$t_c = \begin{cases} t_c^{min} = \frac{d}{v_e+v_h} \sim 100~ps & \text{if } x = \frac{v_e d}{v_e+v_h} \\ t_c^{max} = \frac{d}{v_h} \sim 210~ps & \text{if } x = d \end{cases} \quad (5.1)$$

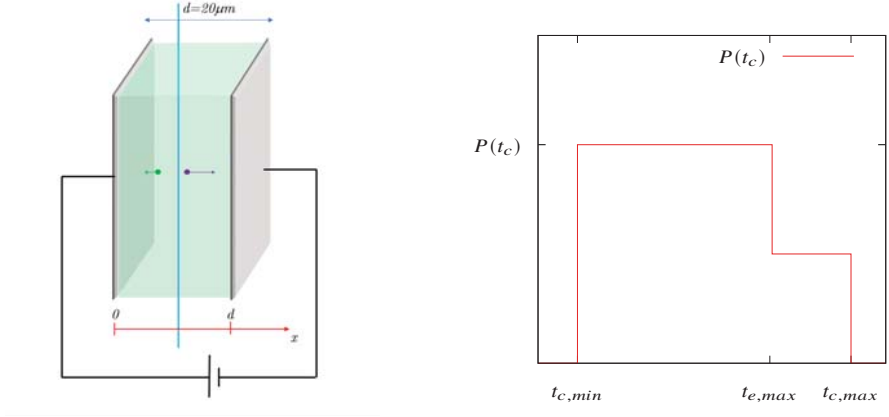

**Figure 16**. Ideal parallel-plate sensor with 3D geometry (left), Charge collection time distribution (right)

For simplicity we can assume that the charge collection times generated by the two carriers are equally probable. In reality, shorter charge collection times are more probable, as electrons move faster (figure 16, right). Following such assumption we will have a rectangular distribution corresponding in the ideal case to a dispersion:

$$\sigma_{t_c} \sim \frac{|t_c^{max} - t_c^{min}|}{\sqrt{12}} \sim 32 ps \quad (5.2)$$

Considering a more realistic description, we can refer to a 3D-trench pixel. Figure 18 shows two distributions obtained by TCoDe simulation [5], referred to two different pixel sizes. In this case, the sensor time behavior and associated time distribution can be characterized by its average collection time instead of a single $t_c$ and by its standard deviation $\sigma_{t_c}$.

## 5.2 Front-end contribution to time jitter

We consider here the jitter contribution coming from the front-end circuit, depending on signal speed and SNR. The time resolution for a given signal can be estimated using the known equation:

$$\sigma_j = \frac{\sigma_V}{\frac{dV}{dt}} \quad (5.3)$$

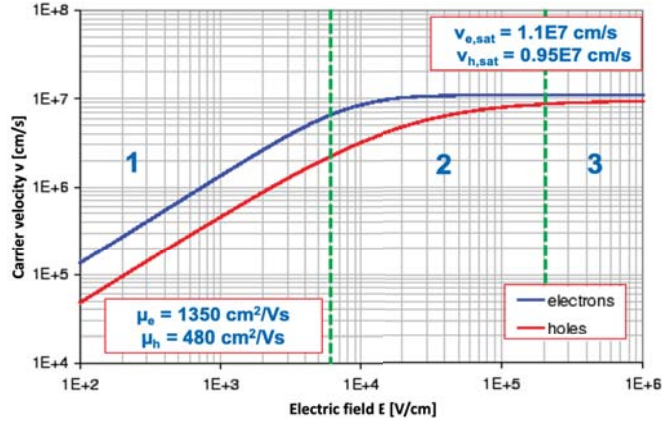

Figure 17. Carrier velocities vs electric field in silicon. 1) low field region; 2) intermediate region; 3) saturation region.

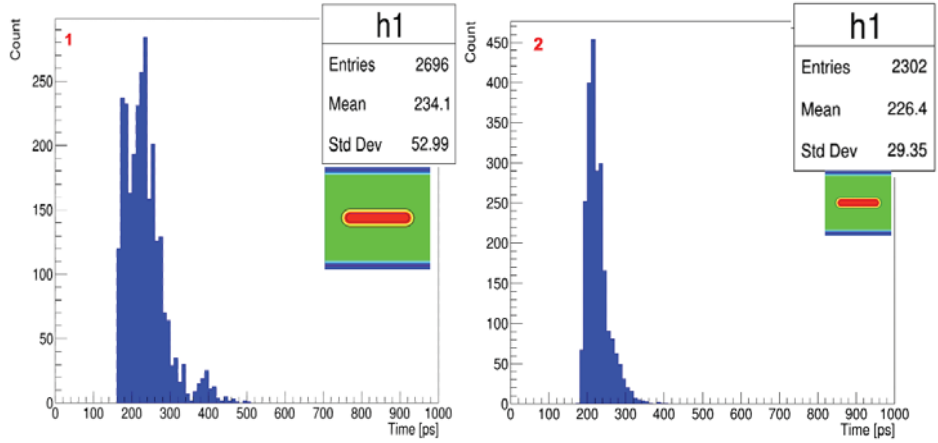

Figure 18. Charge collection time distributions for 3D-trench pixels (TCoDe simulation). Pixel sizes: $55 \times 55 \times 150 \mu m^3$ (left) and $25 \times 25 \times 150 \mu m^3$ (right). The simulated distributions take also into account the pixel dis-uniformities in the electric and weighting fields.

In a time measurement, equation 5.3 is evaluated at a given voltage threshold $V_{th}$. In the present section, exploiting the different expressions calculated in section 4, we analyse the effect of such operation (called *discrimination*) in the two cases of CS-TIA and Fast-TIA. We consider two kind of typical discrimination techniques: the *leading-edge* (*LE*) discriminator and the *constant fraction discriminator* (CFD).

It is important to point out that in what follows the specific amplitude-correction techniques (time-over-threshold and constant fraction discrimination), which are needed to cure the additional dispersion due to *time-walk* (equation 2.1) are considered as already applied independently. In other words we limit our discussion to the intrinsic jitter contribution on the sensor and front-end sides, considering the *time-walk* as a mere systematic (and processing-recoverable) effect.

### 5.2.1 Leading-edge time resolution in CS-TIA

With a leading edge we have a fixed threshold at voltage $V_{th}$. If we set this value as low as possible, the time at which the threshold is crossed $t_{le}$ would be small compared to the time constant of the circuit $\tau$. Assuming this, the exponential term in equation 4.11 can be approximated as:

$$V_{th} = V_{out}(t_{le})_{t>t_c} = Ae^{-\frac{t_{le}}{\tau}}\left(B\frac{t_{le}}{\tau} + C\right) \approx A\left(B\frac{t_{le}}{\tau} + C\right) \tag{5.4}$$

solving for $t_{le}$ we find:

$$t_{le} = \frac{\tau}{B}\left(\frac{V_{th}}{A} - C\right) \tag{5.5}$$

$$t_{le} = \frac{\tau^2(1 - e^{\frac{t_c}{\tau}}) + e^{\frac{t_c}{\tau}}t_c(\tau + \tau_z)}{(e^{\frac{t_c}{\tau}} - 1)(\tau + \tau_z)} + \frac{\tau^2}{(e^{\frac{t_c}{\tau}} - 1)(\tau + \tau_z)}\frac{V_{th}t_c}{Q_{in}R_f}\frac{1 + G_0}{G_0} \tag{5.6}$$

The uncertainty on $t_{le}$ can be found taking the derivative with respect to $t_c$ and propagating the error:

$$\sigma_{t_{le}} = \left(\frac{\partial t_{le}}{\partial t_c}\right)\sigma_{t_c} \tag{5.7}$$

The term $\frac{\partial t_{le,CFD}}{\partial t_c}$ can be defined as the *timing propagation coefficient* $\mathscr{P}$ of time resolution from sensor to electronics. In this specific case we have:

$$\mathscr{P} = \frac{\partial t_{le}}{\partial t_c} = \frac{e^{\frac{t_c}{\tau}}(\tau(e^{\frac{t_c}{\tau}} - 1) - t_c)}{\tau(e^{\frac{t_c}{\tau}} - 1)^2} + \frac{e^{\frac{t_c}{\tau}}(\tau - t_c) - \tau}{\tau(e^{\frac{t_c}{\tau}} - 1)^2}\frac{\tau^2 V_{th}(1 + G_0)}{(\tau + \tau_z)Q_{in}R_f G_0} \tag{5.8}$$

We can consider two contribution to the time at the threshold: the first is independent on $V_{th}$ while the other gets smaller for low value of the threshold. The first term of equation 5.8 becomes $\frac{1}{2}$ for large $\tau$ while the second term has an oblique asymptote of value $-\frac{1}{2}\frac{V_{th}(1+G_0)}{Q_{in}R_f G_0}$ as a function of $\tau$. We can write then:

$$\frac{\partial t_{le}}{\partial t_c} \sim \frac{1}{2}\left(1 - \frac{V_{th}\tau(1 + G_0)}{Q_{in}R_f G_0}\right) \tag{5.9}$$

defining the voltage $V_0 = \frac{Q_{in}R_f}{\tau}$ (with $V_0 \sim e \cdot V_{peak}$) and using $\frac{1+G_0}{G_0} \approx 1$:

$$\frac{\partial t_{le}}{\partial t_c} \sim \frac{1}{2}\left(1 - \frac{V_{th}}{V_0}\right) \tag{5.10}$$

For time constant $\tau \gg t_c$ and low thresholds we have that the uncertainty on the time at the threshold is:

$$\sigma_{t_{le}} \approx \frac{\sigma_{t_c}}{2} \tag{5.11}$$

In the leading-edge configuration, the CS-TIA solution is therefore capable to improve the intrinsic sensor jitter by about a factor 2 (see also [8]): $\mathscr{P} \approx 0.5$.

### 5.2.2 CFD time resolution in CS-TIA

Using a CFD discriminator the threshold is always set at the same fraction of the maximum value of the output voltage. This solution is useful to correct the time walk due to the fact the different amplitudes would cross a fixed threshold at different times. The voltage at the threshold can be written as:

$$V_{th} = \alpha V_{out}^{peak} \approx \alpha \frac{Q_{in} R_f}{e\tau} \tag{5.12}$$

where $\alpha$ define the chosen fraction. Using Eq. 5.8 we find for the time at the threshold in the CFD case:

$$t_{CFD} = \frac{\tau^2(1 - e^{\frac{t_c}{\tau}}) + e^{\frac{t_c}{\tau}} t_c (\tau + \tau_z)}{(e^{\frac{t_c}{\tau}} - 1)(\tau + \tau_z)} + \frac{\alpha}{e} \frac{\tau t_c}{(e^{\frac{t_c}{\tau}} - 1)(\tau + \tau_z)} \tag{5.13}$$

Taking the derivative we find:

$$\frac{\partial t_{CFD}}{\partial t_c} \sim \frac{1}{2}\left(1 + \alpha \frac{2}{e} \frac{\tau}{\tau + \tau_z}\right) \tag{5.14}$$

this approximation holds for low thresholds but also in this case we have a dominant term that is independent of the chosen threshold so that we can estimate the time resolution using:

$$\sigma_{t_{CFD}} = \left(\frac{\partial t_{CFD}}{\partial t_c}\right) \sigma_{t_c} \tag{5.15}$$

$$\sigma_{t_{CFD}} \approx \frac{\sigma_{t_c}}{2} \tag{5.16}$$

Using both *leading-edge* and *CFD*, the time resolution is about half of the standard deviation of the charge collection times distribution. This is a general result when $\tau \gg t_c$ as we will see in section 6

### 5.2.3 Leading-edge time resolution in Fast-TIA

Using the expression of $V_{out}(t)_{t<t_c}$, we can first use an approximation at the second order for the exponential term:

$$V_{out}(t) = I_0 R_{m_0}(1 - e^{-\frac{t}{\tau}}(1 + \frac{t}{\tau})) \tag{5.17}$$

$$V_{th} = I_0 R_{m_0}(1 - e^{-\frac{t_{le}}{\tau}}(1 + \frac{t_{le}}{\tau})) \tag{5.18}$$

$$V_{th} \sim I_0 R_{m_0}(1 - (1 - \frac{t_{le}}{\tau})(1 + \frac{t_{le}}{\tau})) \tag{5.19}$$

$$t_{le} = \tau \sqrt{\frac{V_{th}}{I_0 R_{m_0}}} = \tau \sqrt{\frac{V_{th} t_c}{Q_{in} R_{m_0}}} \tag{5.20}$$

propagating the fluctuations we find:

$$\sigma_{t_{le}} = \frac{t_{le}}{t_c} \frac{\sigma_{t_c}}{2} \tag{5.21}$$

– 27 –

By setting $V_{th}$ to a small value according to a $t_{le} < t_c$ condition, we can reduce the sensor jitter (see equation 4.75 as an example). However, the approximation of equation 5.19 tends to underestimate the fluctuations, being valid only for very small threshold values. For a more complete understanding we can consider the time $t$ as a function of $t_c$ and derive both sides of the equation 5.18 as follows:

$$\frac{\partial}{\partial t_c}\left(\frac{V_{th} t_c}{Q_{in} R_{m_0}}\right) = \frac{\partial}{\partial t_c}\left(1 - e^{-\frac{t_{le}(t_c)}{\tau}}\left(1 + \frac{t_{le}(t_c)}{\tau}\right)\right) \tag{5.22}$$

$$\frac{V_{th}}{Q_{in} R_{m_0}} = \frac{e^{\frac{t_{le}}{\tau}} t_{le}}{\tau^2} \frac{\partial t_{le}}{\partial t_c} \tag{5.23}$$

solving for $\frac{\partial t_{le}}{\partial t_c}$ we can write the derivative as:

$$\frac{\partial t_{le}}{\partial t_c} = \frac{V_{th}}{V'(t_{le}) t_c} \tag{5.24}$$

the resolution at threshold becomes:

$$\sigma_{t_{le}} = \frac{V_{th}}{V'(t_{le}) t_c} \sigma_{t_c} \tag{5.25}$$

Let's suppose to set the threshold to the value corresponding to the maximum slope condition when $t = t_c$. In this case, we can use equations 4.53 and 4.61. The time resolution is then given by:

$$\sigma_{t_{le}} \approx (e - 2)\sigma_{t_c} \sim 0.71 \cdot \sigma_{t_c} \tag{5.26}$$

Setting $V_{th} = V_{\substack{Max \\ Slope}}$ leads to bring about 70% of the charge collection time fluctuations to the time resolution at the threshold. The total time resolution, in this case, would be dominated by the sensor contribution with respect to the front-end one. Setting a lower $V_{th}$ increases the electronic jitter but lowers significantly the sensor contribution. Using the equation 5.18 and the definition of the slope we can write the time resolution as:

$$\sigma_{t_{le}} = \frac{I_0 R_{m_0}\left(1 - e^{-\frac{t_{le}}{\tau}}\left(1 + \frac{t_{le}}{\tau}\right)\right)}{\frac{I_0 R_{m_0} e^{-\frac{t_{le}}{\tau}} t_{le}}{\tau^2} t_c} \sigma_{t_c} \tag{5.27}$$

$$\sigma_{t_{le}} = \left\{\frac{\left(1 - e^{-\frac{t_{le}}{\tau}}\left(1 + \frac{t_{le}}{\tau}\right)\right)}{\frac{e^{-\frac{t_{le}}{\tau}} t_{le}}{\tau^2} t_c}\right\} \sigma_{t_c} \tag{5.28}$$

rearranging equation 5.28 and considering a fixed value for $t_c$ and $\tau$ we can write the following expression as a function of $t_{le}$:

$$\left\{\frac{\partial t_{le}}{\partial t_c}(t_{le})\right\}_{t_{le} < t_c} = \frac{\tau^2(e^{\frac{t_{le}}{\tau}} - 1) - \tau t_{le}}{t_{le} \cdot t_c} \tag{5.29}$$

We will use this expression in section 6, when we will discuss the results obtained in the different discrimination techniques for different front-end solutions.

### 5.2.4 CFD time resolution in Fast-TIA

To find the derivative $\frac{\partial t}{\partial t_c}$ for the constant fraction case we can consider the following equation:

$$V_{out}(t^*)_{t<t_c} = \alpha V_{out}(T_{peak})_{t>t_c} \tag{5.30}$$

$t^*$ is the time at threshold, fixed at the fraction $\alpha$ of the voltage peak $V_{peak} = V_{out}(T_{peak})_{t>t_c}$. In the special case $\tau = t_c$ the peaking time is given by Eq. 4.58, otherwise is given by:

$$T_{peak} = \frac{e^{\frac{t_c}{\tau}} t_c}{e^{\frac{t_c}{\tau}} - 1} \tag{5.31}$$

Eq. 5.30 can be rewritten in the following way:

$$I_0 R_{m_0}(1 - e^{-\frac{t^*}{\tau}}(1 + \frac{t^*}{\tau})) = \alpha I_0 R_{m_0} e^{-\frac{T_{peak}}{\tau}}(e^{\frac{t_c}{\tau}} - 1) \tag{5.32}$$

Taking the derivative with respect to $t_c$ of both side we get:

$$e^{-\frac{t^*}{\tau}} \frac{t^*}{\tau} \frac{\partial t^*}{\partial t_c} = \alpha \frac{\partial}{\partial t_c}\left(e^{-\frac{T_{peak}}{\tau}}(e^{\frac{t_c}{\tau}} - 1)\right) \tag{5.33}$$

the fraction $\alpha$ can be written using Eq. 5.32 as:

$$\alpha = \frac{(1 - e^{-\frac{t^*}{\tau}}(1 + \frac{t^*}{\tau}))}{e^{-\frac{T_{peak}}{\tau}}(e^{\frac{t_c}{\tau}} - 1)} \tag{5.34}$$

the derivative $\frac{\partial t}{\partial t_c}$ is then:

$$\frac{\partial t^*}{\partial t_c} = \alpha \frac{\frac{\partial}{\partial t_c}\left(e^{-\frac{T_{peak}}{\tau}}(e^{\frac{t_c}{\tau}} - 1)\right)}{e^{-\frac{t^*}{\tau}} \frac{t^*}{\tau}} \tag{5.35}$$

$$\frac{\partial t^*}{\partial t_c} = \frac{(1 - e^{-\frac{t^*}{\tau}}(1 + \frac{t^*}{\tau}))}{e^{-\frac{T_{peak}}{\tau}}(e^{\frac{t_c}{\tau}} - 1)} \frac{\frac{\partial}{\partial t_c}\left(e^{-\frac{T_{peak}}{\tau}}(e^{\frac{t_c}{\tau}} - 1)\right)}{e^{-\frac{t^*}{\tau}} \frac{t^*}{\tau}} \tag{5.36}$$

Using the value of $T_{peak}$ (Eq. 5.31) we can write the expression as:

$$\left\{\frac{\partial t_{CFD}}{\partial t_c}(t^*)\right\}_{t^*<t_c} = \frac{t_c}{\tau} \frac{e^{\frac{t_c}{\tau}}(\tau e^{\frac{t^*}{\tau}} - t^* - \tau)}{t^*(e^{\frac{t_c}{\tau}} - 1)^2} \tag{5.37}$$

If we use the same method as eq. 5.30 but considering the signal for $t > t_c$, we find that the derivative $\left\{\frac{\partial t_{CFD}}{\partial t_c}(t^*)\right\}_{t^*>t_c}$ is independent of $t^*$ and can be written:

$$\left\{\frac{\partial t_{CFD}}{\partial t_c}\right\}_{t^*>t_c} = \frac{e^{\frac{t_c}{\tau}}(\tau e^{\frac{t_c}{\tau}} - \tau - t_c)}{\tau(e^{\frac{t_c}{\tau}} - 1)^2} \tag{5.38}$$

which is exaclty the derivative of eq. 5.31:

$$\left\{\frac{\partial t_{CFD}}{\partial t_c}\right\}_{t^*>t_c} = \frac{\partial T_{peak}}{\partial t_c} \tag{5.39}$$

Taking the limit for $\tau \longrightarrow \infty$ (CS-TIA case) of Eq. 5.38 we find:

$$\lim_{\tau \to \infty} \frac{e^{\frac{t_c}{\tau}}(\tau e^{\frac{t_c}{\tau}} - \tau - t_c)}{\tau(e^{\frac{t_c}{\tau}} - 1)^2} = \frac{1}{2} \tag{5.40}$$

solving numerically equation 5.34 for $t^*$ for different $\alpha$, we can find the value of the derivative $\frac{\partial t}{\partial t_c}$ as a function of the threshold.

## 6 Discussion on FB-TIA timing performance and conclusions

Starting from our analysis on the TIA characteristics of the previous sections, in section 5 we have analyzed in detail the effect of time measurement on the two basic TIA scheme: CSA-like, or charge sensitive TIA, and Fast-TIA, which could be also defined as a current-sensitive TIA.

When the dependence of time resolution $\sigma_t$ with respect to the charge collection dispersion $\sigma_{t_c}$ is considered, we have seen that for the CSA-TIA both the CFD and LE discrimination techniques converge to the same value (equations 5.11 and 5.16). In other words, for $\tau \gg t_c$, the front-end electronics is capable to reduce the sensor intrinsic time dispersion up to a factor 2 and the propagation coefficient is $\mathcal{P} \approx 0.5$. Referring to figure 18, this means that a CSA-TIA based front-end could take the time resolution of a suitable 3D-trench sensor to the range of approximately 26 ps (55 $\mu$m pitch) to 15 ps (25 $\mu$m pitch).

However, as already stated at the beginning of this work, the difference between the CSA-TIA and the Fast-TIA is nothing of conceptually fundamental, and is based only on the relative size of the $\tau \sim T_{peak}$ value with respect to the average charge collection time $t_c$ in the sensor (see figures 16, right and 18). This important statement will be better clarified in a moment concerning the circuit behaviour relatively to its timing performances.

On the other hand, when the Fast-TIA case is concerned, the front-end becomes capable to appreciate the shape of the induced signals and the analysis is more complicated and diversified, depending strongly on the threshold position which is possible to choose, according to the noise level of the system. In the simplified case of very low threshold, starting from equation 5.21, we have that:

$$\sigma_{t_{le}} = \frac{\tau}{2t_c}\sqrt{\frac{V_{th}}{I_0 R_{m_0}}}\sigma_{t_c} \approx \frac{\tau}{2t_c}\sqrt{\frac{5\sigma_n}{V_{peak}}}\sigma_{t_c} = \mathcal{P}\sigma_{t_c} \tag{6.1}$$

$$\mathcal{P} \approx \frac{\sqrt{5}\tau}{2t_c}\sqrt{\frac{N}{S}} \qquad (6.2)$$

where we are assuming to place the threshold at 5 times the voltage noise $\sigma_n$. Equations 6.1 and 6.2 show that for $\tau < t_c$ and if a high SNR is obtained, the output from the front-end can reduce considerably the native signal time dispersion of the sensor $\sigma_{t_c}$. As an example, considering a case with $\tau = 0.5\, t_c$, SNR = 30, equation 6.2 would give $\mathcal{P} \approx 10\%$.

More generally, the Fast-TIA behavior can be illustrated in figure 19, left size. This plot is obtained using equations 5.29 and 5.37 and considering induced current signals having rectangular shapes and time width $t_c$, whose time integral corresponds to a given total deposited charge. Here the amplitude fluctuation are not considered, as the signals all correspond to the same charge amount. Therefore the plot can be interpreted as the output performance of the CFD and LE cases, once the time-walk effect is compensated. It can be seen that for a $t_c = 2\tau$ and low threshold, the $\mathcal{P}$ coefficient can be very low, while when the threshold is increased the front-end tends to propagate entirely the sensor jitter contribution to the output and $\mathcal{P}$ approaches unity. The Fast-TIA configuration needs to operate at sufficiently low threshold to be effective on time resolution, otherwise its performance becomes rapidly worse than the CSA-TIA case.

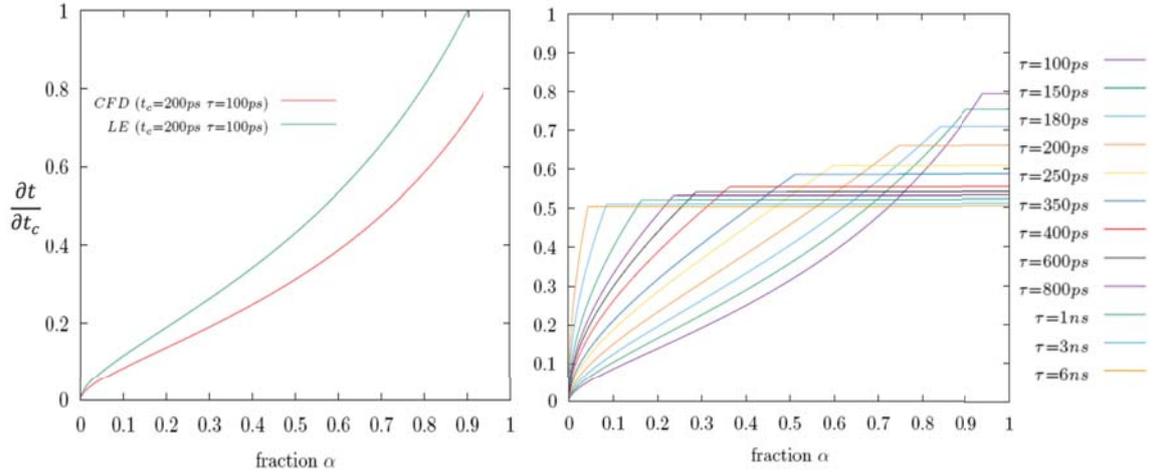

**Figure 19**. Left: derivative $\frac{\partial t}{\partial t_c}$ for fixed $\tau$ and fixed average $t_c = 200\ ps$ as a function of threshold, corresponding to the given fraction $\alpha$ of $V_{peak}$. Right: same plot for CFD with different values of $\tau$.

Figure 19 (right) is a clear demonstration of the fact that the CSA-TIA configuration behaviour is the limit of the Fast-TIA one, when the ratio $\frac{\tau}{t_c}$ increases. A CSA-like behaviour is already observable very early, when $\frac{\tau}{t_c} \approx 2$. This is an independent confirmation of what already seen above (see equation 4.81).

In a circuit having time constant $\tau \sim 1\ ns$, the signal reaches about 20% of $V_{peak}$ at time $t = t_c$, corresponding to already approximately 50% of the charge collection time fluctuations. With a faster front-end electronics, even setting the threshold at a higher fraction, it is still possible to reduce the charge collection time fluctuations to $\approx 20\%$ of the total.

A better estimate of the $\mathcal{P}$ coefficient can be found modeling the current signals of an ideal 3D trenched sensor with charge carriers both at saturation using Ramo theorem. The standard deviation $\sigma_{t_c}$ is given by the charge collection time distribution (Fig. 16 right). In this case, in order to calculate $\mathscr{P}$, a numerical convolution of such current signals with several pulse responses at different time constants $\tau$ can be done. Figure 20 shows the behavior of different front-end stages characterized by different $\tau$ constants: a fast front-end can provide approximately a double reduction of the charge collection time fluctuations. The use of a FAST-TIA configuration allows exploiting deeper and more effectively the performance of intrinsically fast sensors, as for example those realized in 3D technology and in particular in the trenched-electrode geometry.

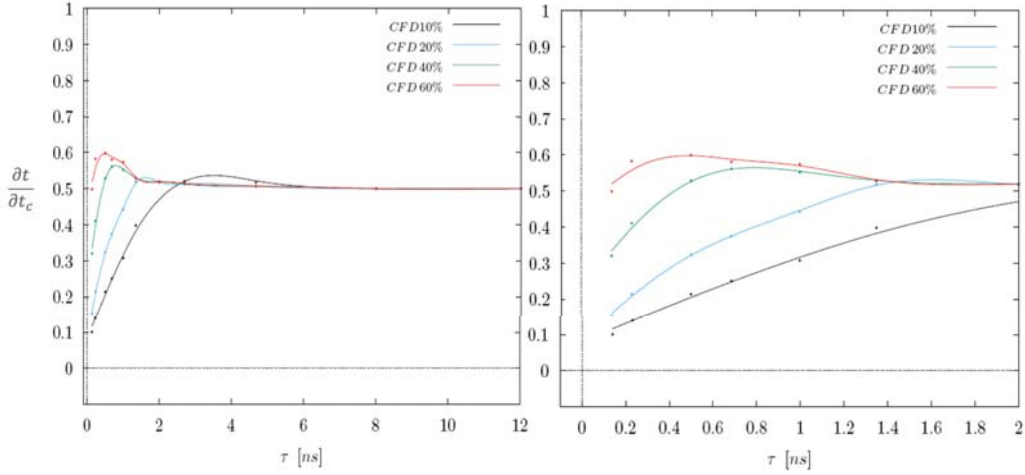

**Figure 20**. Function $\frac{\partial t}{\partial t_c}$ of an ideal 3D detector with electrodes spaced $25\mu m$ for different time constant $\tau$

## A  Appendix A: Calculus of the noise contributions in the CSA-TIA configuration

To calculate the output noise we can consider the input referred noise and use the transfer function of the circuit [7]. The sum in quadrature of the three sources is:

$$I^2_{in,TOT} = I^2_{in,M_1} + I^2_{in,R_D} + I^2_{in,R_f} \tag{A.1}$$

$$I^2_{in,TOT} = \frac{4k_BT\gamma}{g_mR_f^2} + \frac{4k_BT}{g_m^2R_f^2R_D} + \frac{4k_BT}{R_f} \tag{A.2}$$

The transfer function is given by eq. 3.28. The integral of the squared module can be calculated using:

$$R_m(s) = \frac{R_fG_0}{1+G_0}\left(\frac{1}{(1+s\tau)^2} - \frac{s\tau_z}{(1+s\tau)^2}\right) \tag{A.3}$$

$$R_m(f) = \frac{R_fG_0}{1+G_0}\left(\frac{1}{(1+j\frac{f}{f_\tau})^2} - j\frac{f}{f_z(1+j\frac{f}{f_\tau})^2}\right) \tag{A.4}$$

$$|R_m(f)| = \frac{R_fG_0}{1+G_0}\left(\left(\frac{f_\tau^2}{f_\tau^2+f^2}\right)^2 + \left(\frac{ff_\tau^2}{f_z(f_\tau^2+f^2)}\right)^2\right)^{\frac{1}{2}} \tag{A.5}$$

$$\int_0^\infty |R_m(f)|^2 df = \left(\frac{R_f G_0}{1+G_0}\right)^2 \left(\int_0^\infty \frac{f_\tau^4}{(f_\tau^2+f^2)^2} df + \int_0^\infty \frac{f^2 f_\tau^4}{f_z^2(f_\tau^2+f^2)^2} df\right) \quad \text{(A.6)}$$

$$\int_0^\infty |R_m(f)|^2 df = \left(\frac{R_f G_0}{1+G_0}\right)^2 \left(\frac{\pi}{4} f_\tau + \frac{\pi}{4} f_\tau \frac{f_\tau^2}{f_z^2}\right) \quad \text{(A.7)}$$

$$\int_0^\infty |R_m(f)|^2 df = \left(\frac{R_f G_0}{1+G_0}\right)^2 \frac{\pi}{4} f_\tau \left(1 + \left(\frac{\tau_z}{\tau}\right)^2\right) \quad \text{(A.8)}$$

the zero of the transfer function increases the bandwidth of the factor $(\frac{\tau_z}{\tau})^2$ which is normally negligible. The output noise is then given by:

$$\sigma_{n,V_{out}}^2 = I_{in,TOT}^2 \left(\frac{R_f G_0}{1+G_0}\right)^2 \frac{\pi}{4} f_\tau \quad \text{(A.9)}$$

since $\frac{G_0}{1+G_0} \sim 1$ we get:

$$\sigma_{n,V_{out}}^2 \approx \frac{k_B T}{2\tau} \left(\frac{\gamma}{g_m} + \frac{1}{g_m^2 R_D} + R_f\right) \quad \text{(A.10)}$$

On the other hand this method underestimate the contribution given by the two sources $I_{n,M_1}$ and $I_{n,R_D}$. The correct value can be calculated considering the small signal model in figure 21:

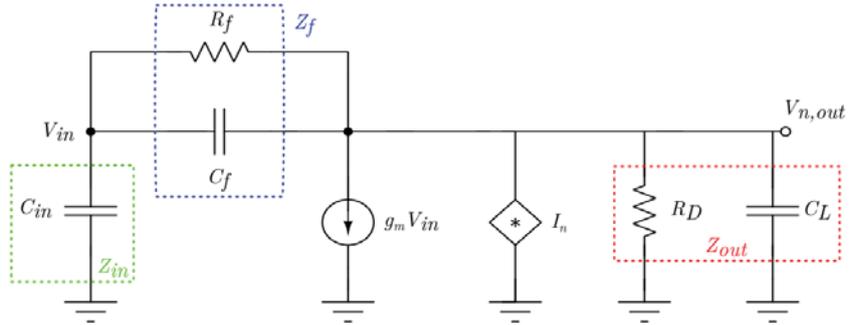

**Figure 21**. Small signal model for the noise sources $I_{n,M_1}$ e $I_{n,R_D}$

Solving the circuit to find $V_{n,out}$ as a function of $I_n$ we get:

$$V_{n,out} = -I_n \frac{Z_{out}}{1 + \frac{(1+g_m Z_{in})Z_{out}}{Z_{in}+Z_f}} \quad \text{(A.11)}$$

substituting the impedances we find the transfer function for the source $I_n$:

$$V_{n,out} = -I_n \frac{R_D(1+sR_f(C_{in}+C_f))}{1+g_m R_D+s(R_f(C_{in}+C_f(1+g_m R_D))+R_D(C_{in}+C_L))+s^2 R_D R_f \xi} \quad (A.12)$$

This transfer function is similar to 3.28 but shows a zero at frequency $f_{z,n} = \frac{1}{2\pi R_f(C_{in}+C_f)}$. It can be shown that multiplying by $\frac{R_f}{R_f+R_D}$ both numerator and denominator we can write eq. A.12 in the following way:

$$V_{n,out} = -\frac{I_n}{g_m} \frac{(1+sR_f(C_{in}+C_f))}{(1+s\tau)^2} \quad (A.13)$$

this expression leads back to the noise found in A.2 for $s = 0$. Considering $M_1$:

$$\sigma^2_{v,M_1} = I^2_{n,M_1} \cdot \frac{1}{g_m^2} \cdot \Delta f = \frac{4k_B T \gamma g_m}{g_m^2} \cdot \Delta f = \frac{4k_B T \gamma}{g_m} \cdot \Delta f \quad (A.14)$$

To find the output noise we use eq. A.8 with $\tau_{z,N} = R_f(C_f + C_{in})$:

$$\sigma^2_{v,M_1} = \frac{I^2_{n,M_1}}{g_m^2} \frac{\pi}{4} f_\tau \left(1 + \left(\frac{\tau_{z,N}}{\tau}\right)^2\right) \quad (A.15)$$

in this case the ratio $\left(\frac{\tau_{z,N}}{\tau}\right)^2$ is bigger than 1 and not negligible:

$$\left(\frac{\tau_{z,N}}{\tau}\right)^2 = \left(\frac{R_f^2(C_{in}+C_f)^2 g_m}{R_f \xi}\right) \quad (A.16)$$

using the definition of $\xi$:

$$\left(\frac{\tau_{z,N}}{\tau}\right)^2 = \frac{g_m R_f (C_f + C_{in})^2}{C_f C_{in} + C_f C_L + C_L C_{in}} \quad (A.17)$$

the dominant term can be written:

$$\sigma_{v,M_1} = \sqrt{\frac{k_B T \gamma}{2}} \frac{C_f + C_{in}}{\xi^{\frac{3}{4}}} g_m^{\frac{1}{4}} R_f^{\frac{1}{4}} \quad (A.18)$$

where $\xi = C_f C_{in} + C_f C_L + C_L C_{in}$. Substituing $I_{n,M_1}$ with $I_{n,R_D}$, we find the noise given by resistor $R_D$:

$$\sigma_{v,R_D} = \sqrt{\frac{k_B T}{2R_D}} \frac{C_f + C_{in}}{g_m^{\frac{1}{4}} \xi^{\frac{3}{4}}} R_f^{\frac{1}{4}} \quad (A.19)$$

while the one given by resistor $R_f$ is:

$$\sigma_{v,R_f} = \sqrt{\frac{k_B T}{2}} \frac{g_m^{\frac{1}{4}}}{R_f^{\frac{3}{4}} \xi^{\frac{1}{4}}} \quad (A.20)$$

## B  Appendix B: Calculus of the noise contributions in the Fast-TIA configuration

The noise introduced by the bipolar transistor is given by the two correlated sources $I_b$ e $I_c$. The PSD's are:

$$I_{n,b}^2 = \frac{2k_B T}{r_\pi}$$

$$I_{n,c}^2 = 2k_B T g_m \tag{B.1}$$

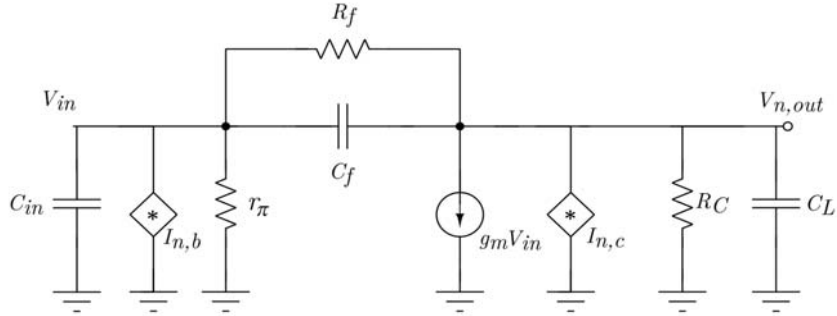

**Figure 22**. Small signal model for the noise sources for the Fast TIA implemented with Bjt

The noise given by source $I_{n,b}$ has to be convoluted with the same transfer function of the signal from the detector, therefore:

$$\sigma_{v,b}^2 = \frac{2k_B T}{r_\pi}(g_m R^* R_{in})^2 \frac{\pi}{4} f_\tau \tag{B.2}$$

$$\sigma_{v,b}^2 = \frac{k_B T}{4 r_\pi \tau}(g_m R^* R_{in})^2 \tag{B.3}$$

To find the noise from source $I_{n,c}$, we can use eq. A.11 and $Z_{in} = r_\pi || C_{in}$ to get the transfer function:

$$V_{n,out} = -I_{n,c} \frac{R_C(r_\pi + R_f)(1 + s R_{\pi f}(C_{in} + C_f))}{R_f + R_C + r_\pi(1 + g_m R_C) + s\left(r_\pi C_{in}(R_f + R_C) + R_f C_f(r_\pi(1 + g_m R_C) + R_C) + R_C C_L(R_f + r_\pi)\right) + s^2 R_C R_f r_\pi \xi} \tag{B.4}$$

where $R_{\pi f} = r_\pi || R_f$.
Simplifying assuming $\zeta = 1$ and using the definition of $\tau$, we get:

$$V_{n,out} = -I_{n,c} \frac{R_C(r_\pi + R_f)}{r_\pi(1 + g_m R_C) + R_C + R_f} \frac{(1 + s\tau_{n,z})}{(1 + s\tau)^2} \tag{B.5}$$

Also in this case we have a zero with time constant:

$$\tau_{n,z} = R_{\pi f}(C_f + C_{in}) \tag{B.6}$$

integrating we get:

$$\sigma_{v,c}^2 = 2k_B T g_m \left( \frac{R_C(r_\pi + R_f)}{r_\pi(1 + g_m R_C) + R_C + R_f} \right)^2 \frac{\pi}{4} f_\tau \left( 1 + \left( \frac{\tau_{n,z}}{\tau} \right)^2 \right) \tag{B.7}$$

using the output resistance of the circuit $R_{out}$ we have:

$$R_{out} = \frac{R_C(r_\pi + R_f)}{r_\pi(1 + g_m R_C) + R_C + R_f} \tag{B.8}$$

$$\sigma_{v,c}^2 = \frac{k_B T g_m R_{out}^2}{4\tau} \left( 1 + \left( \frac{\tau_{n,z}}{\tau} \right)^2 \right) \tag{B.9}$$

the noise from source $I_c$ is given by two contributions:

$$\sigma_{v,c(1)}^2 = \frac{k_B T g_m R_{out}^2}{4\tau} \tag{B.10}$$

$$\sigma_{v,c(2)}^2 = \frac{k_B T g_m R_{out}^2}{4\tau} \left( \frac{\tau_{n,z}}{\tau} \right)^2 \tag{B.11}$$

making clear $\left( \frac{\tau_{n,z}}{\tau} \right)^2$ we can write:

$$\sigma_{v,c(2)}^2 = \frac{k_B T g_m R^* R_{in}(C_f + C_{in})^2}{4\tau \xi} \tag{B.12}$$

we can write it as a single term:

$$\sigma_{v,c}^2 = \frac{k_B T g_m}{4\tau} \left( R_{out}^2 + \frac{R^* R_{in}(C_{in} + C_f)^2}{\xi} \right) \tag{B.13}$$

If $C_{in} \gg C_f$ and $\xi \approx C_{in} C_L$ we get:

$$\sigma_{v,c}^2 \approx \frac{k_B T g_m}{4\tau} \left( R_{out}^2 + R^* R_{in} \frac{C_{in}}{C_L} \right) \tag{B.14}$$

normally the second term is the dominant one. The same analisys can be done for the noise given by resistor $R_C$ leading to:

$$I_{n,c}^2 = 2k_B T g_m \longrightarrow I_{n,R_C}^2 = \frac{4k_B T}{R_C} \qquad (B.15)$$

$$\sigma_{v,R_C}^2 \approx \frac{k_B T}{2\tau R_C}\left(R_{out}^2 + R^* R_{in}\frac{C_{in}}{C_L}\right) \qquad (B.16)$$

For the feedback resistor $R_f$ we can find the transfer function considering the small signal model in figure 23 and the PSD:

$$I_{n,R_f}^2 = \frac{4k_B T}{R_f} \qquad (B.17)$$

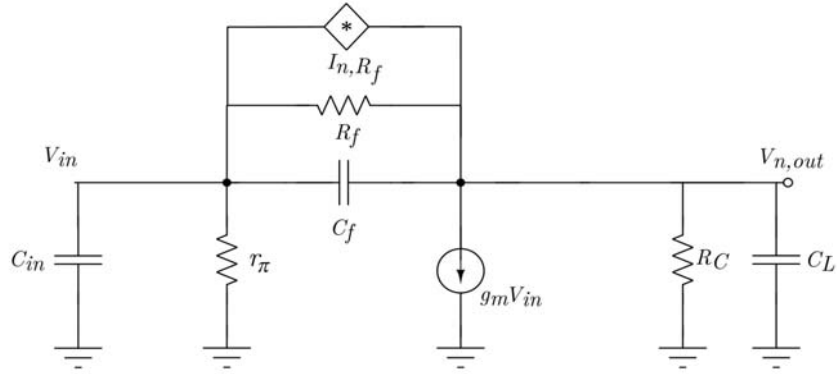

**Figure 23**. Small signal model for resistor $R_f$

$$V_{n,R_f} = I_{n,R_f} \frac{R_f R_C}{R_F + R_C + r_\pi(1+g_m R_C)} \frac{(1 + g_m r_\pi + s r_\pi C_{in})}{(1+s\tau)^2} \qquad (B.18)$$

since $g_m r_\pi \gg 1$ e $g_m R_C \gg 1$ we can consider:

$$V_{n,R_f} = I_{n,R_f} \frac{R_f}{1 + \frac{R_f + R_C}{r_\pi g_m R_C}} \frac{(1 + s\frac{C_{in}}{g_m})}{(1+s\tau)^2} \qquad (B.19)$$

the zero is normally at very high frequencies and we can ignore it:

$$\sigma_{v,R_f}^2 = \frac{k_B T R_f}{(1 + \frac{R_f + R_C}{r_\pi g_m R_C})^2} \frac{1}{2\tau} \qquad (B.20)$$

## Acknowledgments

This work was supported by the Fifth Scientific Commission (CSN5) of the Italian National Institute for Nuclear Physics (INFN), within the Project TIMESPOT. The authors wish to thank Angelo Loi for his help in providing the plots used in figure 18.